\documentclass[superscriptaddress,preprint,nofootinbib]{article}
\pdfoutput=1 
\usepackage{jheppub}
\usepackage{graphicx}
\usepackage{amssymb}
\usepackage{amsmath}
\usepackage{epstopdf}
\usepackage{subfigure}
\usepackage{color}

\graphicspath{{./Artwork/}}

\def\Ab{\bar{A}}
\def\wb{\bar{w}}
\def\be{\begin{equation}}
\def\ee{\end{equation}}
\newcommand\sbraket[1]{\langle #1 \rangle}
\newcommand\braket[1]{[ #1 ]}
\newcommand\slashed[1]{#1\hskip-0.55em /}

\begin{document}

\title{Colour-Kinematics Duality for One-Loop Rational Amplitudes}

\author{Rutger H. Boels${}^a$,}
\author{Reinke Sven Isermann${}^a$,}
\affiliation{${}^a$II. Institut f\"ur Theoretische Physik, Universit\"at Hamburg, Luruper Chaussee 149, D- 22761 Hamburg, Germany}
\author{Ricardo Monteiro${}^b$,}
\author{Donal O'Connell${}^b$}
\affiliation{${}^b$The Niels Bohr International Academy and Discovery Center, The Niels Bohr Institute, Blegdamsvej 17, Copenhagen 2100, Denmark}

\abstract{
\noindent Colour-kinematics duality is the conjecture of a group theory-like structure for the kinematic dependence of scattering amplitudes in gauge theory and gravity. This structure has been verified at tree level in various ways, but similar progress to all multiplicity has been lacking at loop level, where the power of the duality would be most significant. Here we explore colour-kinematics duality at one loop using the self-dual sector as a starting point. The duality is shown to exist in pure Yang-Mills theory for two infinite classes of amplitudes: amplitudes with any number of particles either all of the same helicity or with one particle helicity opposite the rest. We provide a simple Lagrangian-based argument in favour of the double copy relation between gauge theory and gravity amplitudes in these classes, and provide some explicit examples. We further discuss aspects of the duality which persist after integration, leading to relations among partial amplitudes. Finally, we describe form factors in the self-dual theory at tree level which also satisfy the duality.
}

\maketitle
\tableofcontents

\section{Introduction}

Colour-kinematics duality has proven to be an inspirational idea in the study of perturbative gauge theory and gravity. The idea, due to Bern, Carrasco, and Johansson (BCJ), was first introduced at tree level~\cite{Bern:2008qj} before being extended to loop amplitudes~\cite{Bern:2010ue}. In all cases, the principle objects of study are the kinematic numerators of cubic Feynman-like diagrams. Colour-kinematics duality states that whenever the natural colour factors of three cubic diagrams satisfy a Jacobi relation the corresponding numerators can be put in a \textit{dual} form in which they satisfy the same Jacobi relation. A separate, but closely related, idea due to Bern, Carrasco and Johansson relates Yang-Mills amplitudes expressed in a colour dual form to gravitational amplitudes. This conjecture, known as the double copy formula, states that gravity amplitudes can be obtained from the cubic diagrams as a double copy, i.e. by replacing the colour factors by another copy of the numerators. This can be thought of as a generalisation of the famous Kawai-Lewellen-Tye (KLT) relations~\cite{Kawai:1985xq} of tree level string theory which relate closed string amplitudes to a sum over products of open string amplitudes.

At tree level, much is known about the duality and the associated double copy formula. Various authors have described how to compute numerators obeying the duality (we shall call such numerators dual numerators for simplicity) for any tree amplitude in gauge theory~\cite{BjerrumBohr:2010hn,Mafra:2011kj,BjerrumBohr:2012mg,Fu:2012uy}. Similarly, the double copy formula has been proven at tree level using recursive arguments~\cite{Bern:2010yg}. At loop level less is known. As we shall review below, there are several impressive examples of the duality at work for low multiplicity, but it still remains to be understood whether sets of dual numerators can always be found.

The existence of colour-dual numerators implies relations between colour-ordered amplitudes at tree level~\cite{Bern:2008qj}, known as the BCJ relations. These relations have been proven by a variety of different methods~\cite{BjerrumBohr:2009rd,Stieberger:2009hq,Feng:2010my,Cachazo:2012uq}. In principle, similar relations should also exist at loop level if colour-dual numerators exist. There have been several cases in which insights from string theory have been important in studying colour-kinematics duality. The first proofs of the BCJ relations used string methods~\cite{BjerrumBohr:2009rd,Stieberger:2009hq}, and more recently the pure spinor approach to the superstring has been elegantly used to provide insight into the problem of finding dual numerators at tree and one-loop level~\cite{Mafra:2011kj,Mafra:2012kh}.

It is fascinating that colour-kinematics duality hints that there is some kind of algebraic structure underlying gauge theory numerators. There has been progress in identifying this structure~\cite{Monteiro:2011pc} in a simplified case, namely the self-dual sectors of Yang-Mills theory and of gravity. In this simpler setting, it is possible to compute numerators by linking together the structure constants of a certain diffeomorphism algebra. Our work in this paper is, in a sense, a continuation of the work of~\cite{Monteiro:2011pc} by a subset of the present authors. We will examine to what extent the diffeomorphisms allow us to understand loop amplitudes. Indeed, it is known~\cite{Cangemi:1996rx,Chalmers:1996rq} that one-loop amplitudes in self-dual Yang-Mills theory are the same as the all-plus\footnote{The all-minus one-loop amplitudes can be obtained from the anti-self-dual theory, and are simple parity conjugates of the all-plus amplitudes.} one-loop amplitudes of (pure) Yang-Mills theory. Thus, we are easily able to compute dual numerators for an infinite class of one-loop amplitudes. These amplitudes are quite special: the fully integrated amplitudes contain no logarithms in four dimensions; this reflects the fact that cutting any of the one-loop amplitudes in four dimensions leads to a vanishing product of tree amplitudes. In other words, the all-plus amplitudes are rational functions, and in fact they are not the only amplitudes in Yang-Mills theory which are simple rational functions of the external data. The other class are the ``one-minus" amplitudes, in which all particles but one have positive helicity\footnote{The one-plus amplitudes are parity conjugates of the one-minus amplitudes.}. We show below that one can build on our understanding of the self-dual theory to compute dual numerators for the one-minus amplitudes. Thus, the main result of our work is the identification of two infinite classes of amplitudes at one-loop for which colour-kinematics duality does indeed hold.

Given numerators which satisfy colour-kinematics duality, it is natural to use the double copy to compute gravitational amplitudes. We provide a simple argument in support of the double copy relating the finite amplitudes under study here, based on the known light-cone Lagrangians in gauge and gravity theory. In addition, we explicitly check the relation for the all-plus and one-minus four point amplitudes.

The fact that the families of one-loop amplitudes which we deal with are especially simple makes them an ideal laboratory for exploring what happens to colour-kinematics duality after integration. Indeed, it is known that these amplitudes satisfy a set of relations~\cite{BjerrumBohr:2011xe}. We explore whether the integrated amplitude itself can be thought of as being built from some objects which follow naturally from the kinematic algebra, focussing on the simplest case of the all-plus amplitudes. The answer to our question is a qualified yes, as we shall discuss below.

One of the prime motivations for our interest in self-dual Yang-Mills theory is that it makes colour-kinematics duality manifest at the level of the action. This implies that observables calculated in this theory may exhibit the duality. The all-plus amplitudes at one loop are examples of this. The one leg off-shell tree level current is another, but this is gauge variant. Motivated by \cite{Boels:2012ew} we point out that a certain tree level form factor involving the trace of the anti-self-dual field strength tensor squared and only positive helicity on-shell gluons can be calculated within the self-dual Yang-Mills theory in a manifestly colour-dual manner. 

The structure of this paper is as follows. In section~\ref{sec:review} we review colour-kinematics duality in more detail, and discuss the duality in the self-dual sector. We then open the discussion of one-loop numerators for all-plus and one-minus amplitudes in section~\ref{sec:oneloop}. In section~\ref{sec:integrated} we turn to the relations that the integrated all-plus amplitudes satisfy, and explain in what sense they are related to the duality. We turn to the topic of form factors in section~\ref{sec:ffs} before discussing our results in section~\ref{sec:dc}.

%%%%%%%%%%%%%%%%%%%%%%%%%%%%%%%%%%%%%
%%%%%%%%%%%%%%%%%%%%%%%%%%%%%%%%%%%%%

\section{Review and formalism}
\label{sec:review}

We open with a short review to set the stage for our work. First, we review BCJ colour-kinematics duality and the associated double copy formula briefly, to remind the reader of the concepts which will be of principle importance in this article and also to establish some notation. Then we move on to review the duality in the context of the self-dual sectors of Yang-Mills theory and of gravity, and the closely related MHV amplitudes, where it is possible to understand the group theoretic structure of the kinematic dependence in detail. 

\subsection{Generalities of colour-kinematics duality}

For much of this paper, the objects of central concern will be a set of kinematic numerators. Each numerator is associated to a certain cubic diagram. The diagrams at $L$ loops consist of all connected diagrams, with cubic vertices, and the appropriate number of external lines. Given such a graph, the associated colour factor is trivial to write down: a factor $f^{abc}$ is associated with each vertex, and the internal lines receive a factor $\delta^{ab}$. The $L$-loop Yang-Mills amplitude can be written as
\begin{equation}
\label{eq:defAmp}
\mathcal{A}_n^{(L)} = i^L g^{n-2+2L} \sum_{\mathrm{diagrams} \; \alpha}\int \prod_{i=1}^L \frac{d^D l_i}{(2 \pi)^D} \frac{1}{S_\alpha} \frac{c_\alpha n_\alpha(l_i)}{D_\alpha(l_i)},
\end{equation}
where the summation runs over the distinct cubic diagrams with $L$ loops. Under the integral sign are the usual symmetry factors $S_\alpha$, the colour factors $c_\alpha$, the kinematic numerators $n_\alpha$ which may depend on the loop momenta, and canonical scalar-type Feynman propagators which we have combined into a denominator $D_\alpha$.

The colour factors are built out of structure constants of the Lie group underlying the gauge theory. As such, there are many triplets $(\alpha, \beta, \gamma)$ of diagrams such that the Jacobi relation holds among the colour factors:
\begin{equation}
c_\alpha + c_\beta + c_\gamma = 0.
\end{equation}
Let us call a triplet of diagrams with this property a Jacobi triplet. Colour-kinematics duality~\cite{Bern:2008qj,Bern:2010ue} is the assertion that one can always find a valid set of numerators which have the property that, for all Jacobi triplets $(\alpha, \beta, \gamma)$, the numerators satisfy
\begin{equation}
n_\alpha + n_\beta + n_\gamma = 0.
\end{equation}
A valid set of numerators, of course, is simply a set of numerators such that the amplitude in Eq.~\eqref{eq:defAmp} is the correct Yang-Mills amplitude. So the challenge is to find such a set of dual numerators. At tree level, we know~\cite{BjerrumBohr:2010hn,Mafra:2011kj,BjerrumBohr:2012mg,Fu:2012uy,Broedel:2011pd} that such numerators can be found. Beyond tree level, our information is more sporadic. The four-point amplitude in $\mathcal{N} = 4$ super-Yang-Mills (sYM) has been put into a form with numerators satisfying the duality up to four loops~\cite{Bern:2010ue,Bern:1998ug,Bern:2012uf}. The five-point one- and two-loop amplitudes of $\mathcal{N} = 4$ sYM have also appeared~\cite{Carrasco:2011mn} in a manifestly dual form in the literature; see also~\cite{Yuan:2012rg}. The four-point amplitude has also been shown to admit a colour-kinematics dual form in reduced supersymmetry theories at one-loop \cite{Carrasco:2012ca}; in the case of pure Yang-Mills, up to two loops when all helicities are equal \cite{Bern:2010ue}. Recently the two-point form factor was obtained in colour-dual form up to four loops, and the three-point up to two loops \cite{Boels:2012ew}. 

Note that the challenge of finding colour-dual numerators is very closely related to a freedom in shifting the numerators. This freedom is usually called generalised gauge invariance. At tree level this freedom states that a set of numerators can be shifted without changing the amplitude by some amount $\Delta$ if the following holds
\begin{equation}
\label{eq:gengauge}
n_\alpha \rightarrow n'_\alpha = n_\alpha +\Delta_\alpha \quad \text{with} \quad \sum_\alpha \frac{c_\alpha \Delta_\alpha}{D_\alpha}=0
\end{equation}
In addition, if the numerators satisfy the Jacobi identities, then the $\Delta_{\alpha}$ should satisfy the Jacobi relation for each Jacobi triplet. At the integrand level similar formulae can be written down. 

Closely related to the colour-kinematics duality is the double copy conjecture~\cite{Bern:2008qj,Bern:2010ue}. The double copy expresses the close relationship between gauge theory and gravity---in some sense, gravity is the square of gauge theory. Given a dual set of numerators, the double copy conjecture states that a gravity amplitude can be computed as
\begin{equation}
\label{eq:double}
\mathcal{M}_n^{(L)} = i^{L+1} \left(\frac{\kappa}{2}\right)^{n-2+2L} \sum_{\mathrm{diagrams} \; \alpha}\int \prod_{i=1}^L \frac{d^D l_i}{(2 \pi)^D} \frac{1}{S_\alpha} \frac{n_\alpha(l_i) \tilde n_\alpha(l_i)}{D_\alpha(l_i)}.
\end{equation}
Comparing to our expression for a gauge theory amplitude, Eq.~\eqref{eq:defAmp}, the double copy formula simply replaces a colour factor in the gauge theory amplitude with another copy of a kinematic numerator. 
Note that we have two distinct numerators in the double copy formula, Eq.~\eqref{eq:double}, as these numerators need not be computed in the same gauge theory. For example, the $n_\alpha$ could be computed in $\mathcal{N} = 4$ sYM, while the $\tilde n_\alpha$ could be computed in $\mathcal{N} = 0$ sYM (that is, pure Yang-Mills theory). The states being scattered in the gravitational amplitude $\mathcal{M}$ are the outer product of the states being scattered in the gauge theory amplitudes. Our example for $n_\alpha$ and $\tilde n_\alpha$ gives an amplitude of $\mathcal{N} = 4$ supergravity; see Ref.~\cite{Damgaard:2012fb} for a complete map between gauge theories and supergravities. It is expected that only one set of numerators must satisfy colour-kinematics duality, while the other set can be any valid numerators---for instance, those computed directly by Feynman diagrams. This has only generically been proven at tree level~\cite{Bern:2010yg} and checked in various supergravity examples \cite{Bern:2011rj,BoucherVeronneau:2011qv,Bern:2012cd,Carrasco:2012ca}.

The double copy formula has been proven at tree level~\cite{Bern:2010yg}. Beyond tree level, the formula has been checked in several non-trivial cases. Using the known $\mathcal{N}=4$ sYM amplitudes twice, the four-point scattering amplitude in $\mathcal{N}=8$ supergravity has been computed up to four loops~\cite{Bern:2010ue,Bern:1998ug,Bern:2012uf}. Similarly, the five-point $\mathcal{N}=8$ amplitude has been constructed via the double copy~\cite{Carrasco:2011mn} at one and two loops. In an interesting development, the $\mathcal{N} = 8$ six-point amplitude has recently been constructed via the double copy at one loop, without explicitly finding a dual set of numerators~\cite{Yuan:2012rg}. The double copy has also been explored in the context of orbifold theories \cite{Carrasco:2012ca}.

The most powerful aspect of the double copy formula is that, more than a tool, it seems to be crucial for the ultraviolet behaviour of supergravity theories. Ref.~\cite{Bern:2012gh} analysed in detail the consequences for half-maximal supergravity. Very recently, Ref.~\cite{Boels:2012sy} presented a general argument showing how the precise implementation of the colour-kinematics duality at loop level is fundamental for the degree of divergence. Evidence for the all-loop validity of the colour-kinematics duality and the double copy appears in the soft limit \cite{Oxburgh:2012zr}, in some high-energy limits \cite{Saotome:2012vy}, and also in the BCFW shifts of gauge theory integrands \cite{Boels:2012sy}. Remarkably, extensions of these ideas seem to apply to amplitudes in the ABJM theory \cite{Bargheer:2012gv,Huang:2012wr}. The introduction of higher-dimension operators in Yang-Mills theory has also been studied \cite{Broedel:2012rc}.

\subsection{Colour-kinematics duality in the self-dual sector and MHV amplitudes}
\label{sec:sdTree}

The colour-kinematics duality manifests itself naturally in the self-dual sector of gauge theory, and in the closely related MHV sector, as shown in Ref.~\cite{Monteiro:2011pc}. We will review here those results, with a new presentation better adapted to the spinor-helicity formalism, which will be useful later.

Let us recall the form of the Yang-Mills Lagrangian written in light-cone gauge \cite{Chalmers:1998jb}:
\begin{equation}
\label{CSlag}
{\mathcal L}= \textrm{tr}\Big\{\;  \frac{1}{2} \Ab \,\partial^2 A -ig \Big(\frac{\partial_w}{\partial_u}A\Big) [A,\partial_u \Ab] 
- ig \Big(\frac{\partial_{\wb}}{\partial_u}\Ab\Big) [\Ab,\partial_u A] -g^2 [A,\partial_u \Ab]\frac{1}{\partial_u^2} [\Ab,\partial_u A] \;\Big\}.
\end{equation}
The convention here is that the field $A$ carries positive helicity, while $\bar A$ carries negative helicity. The indices correspond to the coordinates
\begin{equation}\label{eq:light-conecoords}
u = t-z, \quad v = t + z, \quad w = x + i y, \quad \bar w = x- i y ,
\end{equation}
and we defined $\partial^2 \equiv 2(\partial_u \partial_v - \partial_w \partial_{\bar w})$. The light-cone condition is $A_u=0$. The Lagrangian has three types of vertices: the $(++-)$ vertex, the $(+--)$ vertex, and the four-point vertex $(++--)$. For instance, the momentum space Feynman rule for the $(++-)$ vertex gives
\begin{equation}
\label{LCvertexppp}
\begin{minipage}[c]{0.3\linewidth}
\centering
\includegraphics[scale=0.4]{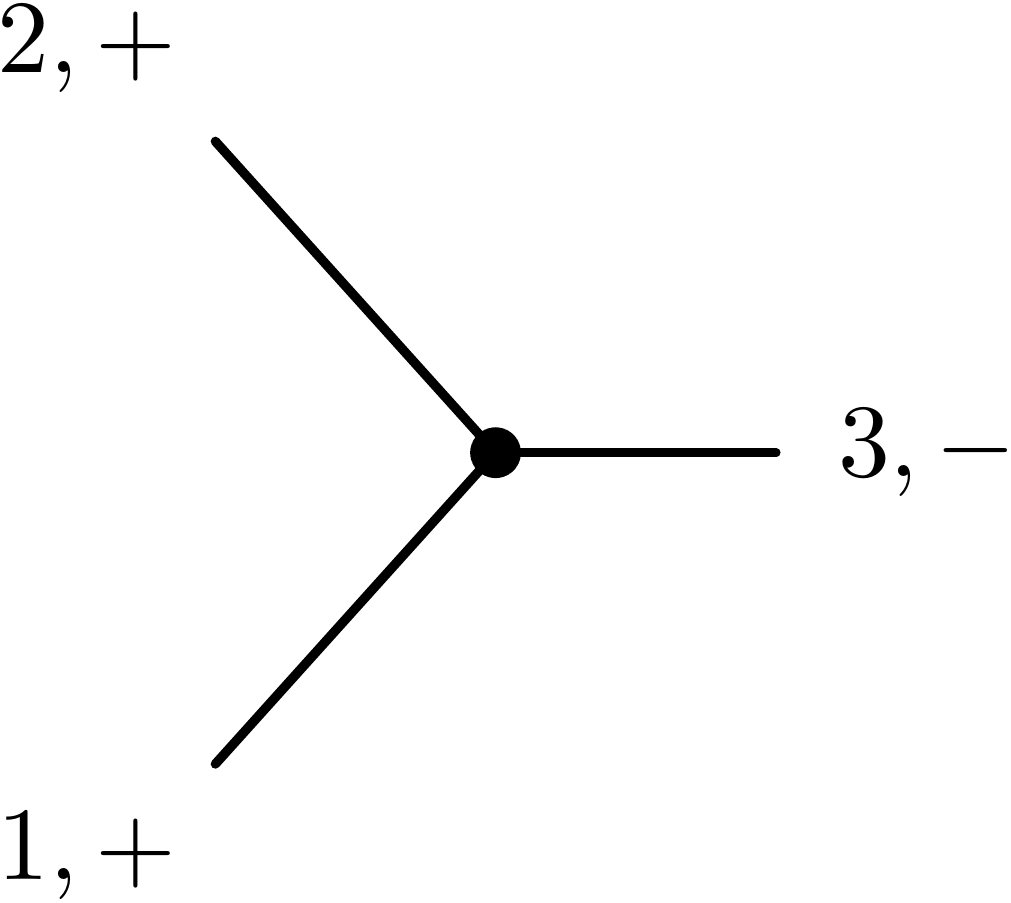}
\end{minipage} = g \frac{p_{3u}}{p_{1u} p_{2u}} (p_{1 w} p_{2u} - p_{2w} p_{1u}) \, f^{a_1 a_2 a_3}.
\end{equation}

It is more convenient for our purposes to write the Feynman rules for the Lagrangian \eqref{CSlag} in terms of the spinor-helicity formalism, for which we now give a concise review; see \cite{Dixon:1996wi} for more background. The basis of the formalism is that any on-shell momentum, $p^2=0$, can be expressed in terms of two spinors, $\lambda$ and $\tilde \lambda$, in the following manner,
\begin{equation}
p_{\alpha \dot{\alpha}} \equiv p_\mu \sigma^\mu_{\alpha \dot{\alpha}} = \lambda_\alpha \lambda_{\dot{\alpha}},
\end{equation}
where $\sigma^\mu=(1_{2\times 2},\sigma^i)$, the $\sigma^i$ representing the Pauli matrices. We can define $SL(2)$-invariant inner products of the spinors between two on-shell particles, say $i$ and $j$,
\begin{equation}
\langle ij \rangle = \epsilon^{\alpha \beta} \lambda_\alpha^{(i)} \lambda_\beta^{(j)} \qquad \textrm{and}  \qquad 
[ ij ] = -\epsilon^{\dot{\alpha} \dot{\beta}} \tilde \lambda_{\dot{\alpha}}^{(i)} \tilde \lambda_{\dot{\beta}}^{(j)},
\end{equation}
such that $(p_i+p_j)^2=\langle ij \rangle [ji]$. This notation can be extended to define
\begin{equation}
\langle i|j|k] = -\lambda_\alpha^{(i)} \epsilon^{\alpha \beta} \,p_{\beta \dot{\alpha}}^{(j)}\, \epsilon^{\dot{\alpha} \dot{\beta}} \lambda_{\dot{\beta}}^{(k)} = [k| j|i \rangle ,
\end{equation}
where $p^{(j)}$ need not be on-shell. For on-shell momenta, we may denote $p=|p \rangle [p|$ to facilitate the inner products. The last ingredients that we need for the purposes of this paper are
\begin{align}
\label{eq:sandwhich}
\langle i|jk|l \rangle &= -\lambda_\alpha^{(i)} \epsilon^{\alpha \beta} \,p_{\beta \dot{\alpha}}^{(j)}\, \epsilon^{\dot{\alpha} \dot{\beta}} \,p_{\gamma \dot{\beta}}^{(k)}\, \epsilon^{\gamma \delta} \lambda_\delta^{(l)} = - \langle l|kj|i \rangle , \nonumber \\
[ i|jk|l ] &= \lambda_{\dot{\alpha}}^{(i)} \epsilon^{\dot{\alpha} \dot{\beta}} \,p_{\alpha \dot{\beta}}^{(j)}\, \epsilon^{\alpha \beta} \,p_{\beta \dot{\gamma}}^{(k)}\, \epsilon^{\dot{\gamma} \dot{\delta}} \lambda_{\dot{\delta}}^{(l)} = - [ l|kj|i ] .
\end{align}
These are natural definitions which, in case $p^{(j)}$ and $p^{(k)}$ are on-shell, lead to
\begin{equation}
\langle i|jk|l \rangle = \langle ij \rangle [jk] \langle kl \rangle , \qquad [ i|jk|l ] = [ ij ] \langle jk \rangle [kl].
\end{equation}

With the help of the spinor-helicity formalism, we define the light-cone with a null vector $\eta=|\eta \rangle [\eta|$, such that $\eta \cdot A = 0$. Omitting the coupling constant, the Feynman rules for the vertices are
\begin{equation}\label{LCvertices}
\begin{gathered}
(i^+,j^+,k^-)=\frac{k_\eta}{i_\eta j_\eta}\, X(i,j) \, f^{a_i a_j a_k}, \qquad \textrm{with} \quad X(i,j) \equiv \langle\eta|ij|\eta\rangle,\\
(i^-,j^-,k^+)=\frac{k_\eta}{i_\eta j_\eta}\, \overline{X}(i,j) \, f^{a_i a_j a_k}, \qquad \textrm{with} \quad \overline{X}(i,j) \equiv [\eta|ij|\eta],\\
(i^+,j^+,k^-,l^-)=i \left( \frac{i_\eta k_\eta + j_\eta l_\eta}{(i_\eta+l_\eta)^2} \, f^{a_i a_l b} f^{b a_j a_k}
+ \frac{i_\eta l_\eta + j_\eta k_\eta}{(i_\eta+k_\eta)^2} \, f^{a_i a_k b} f^{b a_j a_l} \right),
\end{gathered}
\end{equation}
where we defined $i_\eta =\langle\eta|i|\eta]$. Each propagator contributes as
\begin{equation}
\frac{i}{(p_i+p_j)^2}\,\delta^{a_i a_j} ,
\end{equation}
and the polarization factor for each external particle is
\begin{equation}\label{LChel}
e_i^{(+)}=\frac{[\eta i]}{\langle \eta i\rangle}, \qquad e_i^{(-)}=\frac{\langle \eta i\rangle} {[\eta i]}.
\end{equation}
Notice that $X$ and $\overline X$ are antisymmetric by virtue of Eq.~\eqref{eq:sandwhich}. They correspond to spinor products, for instance $X(i,j)=-[\hat{i},\hat{j}]$, where the `hat' spinors are defined from a possibly off-shell momentum as $|\hat{i}]_{\dot{\alpha}}=p^{(i)}_{\alpha \dot{\alpha}} \epsilon^{\alpha \beta} |\eta\rangle_\beta$.
The introduction of the spinors $|\eta \rangle$ and $|\eta ]$ makes the freedom in choosing the light-cone direction manifest. Scattering amplitudes are invariant for the choice of these spinors. The rule \eqref{LCvertexppp} for gluons is recovered with the choice $|\eta \rangle \sim (1,0)^T$ and $\langle \eta | \sim (1,0)$.

As a first example of this notation and as a consistency check, it is instructive to derive the three-point $ \overline{\text{MHV}}$ amplitude from the above rules,
\begin{equation}
 e_i^{(+)}e_j^{(+)}e_k^{(-)}\frac{k_\eta}{i_\eta j_\eta}X(i,j) = \frac{[\eta i]}{\langle \eta i \rangle}\frac{[\eta j]}{\langle \eta j \rangle}\frac{\langle \eta k \rangle}{[\eta k]}\frac{\langle \eta  k \rangle [k\eta]}{\langle \eta i \rangle [i\eta]\langle \eta  j \rangle [j\eta]}\langle \eta i \rangle [i j]\langle j \eta \rangle=\frac{[ij]^3}{[jk][ki]}.
\end{equation}
We can also consider a minimally coupled scalar. The scalar-gluon-scalar vertex is
\begin{equation}
\label{SGSvertices}
(i_s,j^+,k_s)=\frac{1}{j_\eta}X(i,j) ,  \qquad  (i_s,j^-,k_s)= \frac{1}{j_\eta}\overline X(i,j),
\end{equation}
so that the three-point partial amplitude (where the scalars may be massive) reads
\begin{equation}
A_3(i_s,j^+,k_s)=e_j^{(+)}\frac{1}{j_\eta}X(i,j) = - \frac{X(i,j)}{\langle \eta j \rangle^2} = \frac{\langle \eta|i|j]}{\langle \eta j \rangle}, \qquad 
A_3(i_s,j^-,k_s)=\frac{[ \eta|i|j\rangle}{[ \eta j ]}.
\end{equation}

The self-dual sector of gauge theory is the restriction of the full Yang-Mills theory to the vertex $(++-)$. Therefore, there are only cubic vertices in the relevant Feynman diagrams, which is the first step in order to have a manifest colour-kinematics dual form. The other requirement is that, whenever the colour factors of three diagrams satisfy Jacobi identities, 
\begin{equation}
\label{fjacobi}
f^{a_i a_j b} f^{b a_k a_l} + f^{a_j a_k b} f^{b a_i a_l} + f^{a_k a_i b} f^{b a_j a_l} = 0,
\end{equation}
the kinematic numerators of those diagrams satisfy the same identities. A brief inspection of the $(++-)$ vertex in \eqref{LCvertices} shows that this requirement holds if we have
\begin{equation}
\label{Xjacobi}
X(i,j) X(k,l) + X(j,k) X(i,l) + X(k,i) X(j,l) =0.
\end{equation}
This is a consequence of the Schouten identity for the spinors defined as $|\hat p]= p|\eta\rangle$. As shown in Ref.~\cite{Monteiro:2011pc}, it can also be seen as the Jacobi identity for the Lie algebra of area-preserving diffeomorphisms. So the colour-kinematics duality is manifest in the self-dual sector.

The only tree-level Feynman diagrams in the self-dual sector have external helicities $-++\ldots +$. The ``one-minus" scattering amplitudes obtained from those diagrams are known to vanish (except the three-point amplitude for complex momenta). However, MHV amplitudes, which have helicity structure $--+\ldots +$, are closely related. A simple counting argument\footnote{We will present a similar argument in the next section in the context of one-loop amplitudes.} shows that the Feynman diagrams contributing to MHV amplitudes have only one vertex which is not of the type $(++-)$. This could be a $(--+)$ vertex or a four-point vertex. Using the freedom to choose the null vector $\eta$ defining the light-cone, it is possible to eliminate the diagrams with a four-point vertex.

The procedure for MHV amplitudes is as follows. Consider one of the two negative helicity particles, say particle $1$. Then take the limit
\begin{equation}
\label{MHVlimit}
|\eta\rangle \to |1\rangle.
\end{equation}
In this limit, we have $e_1^{(-)} \to 0$. However, we also have $1_\eta \to 0$. What happens in this limit is that all diagrams which do not have a pole in $1_\eta$ vanish. An inspection of the Feynman rules \eqref{LCvertices} shows that such a pole is only possible if particle 1 is attached to a $(--+)$ vertex. Therefore, the MHV amplitudes in this gauge contain graphs with only cubic vertices, one of them of the type $(--+)$, and all the others of the type $(++-)$. For colour-kinematics duality to be manifest, we need to have Jacobi-like identities. Most of these will involve only $(++-)$ vertices, so the story is the same as for the self-dual sector. When the identities involve the single $(--+)$ vertex (that is, when they involve the external leg 1), we have, instead of \eqref{Xjacobi}, the requirement
\begin{equation}
\label{MHVjacobi}
\overline X(1,i) X(j,k) + \overline X(1,j) X(k,i) + \overline X(1,k) X(i,j) =0.
\end{equation}
Again, this is just a consequence of the Schouten identity, as we can see by rewriting the left-hand-side as
\begin{equation}
\label{MHVshouten}
[\eta 1] \big(  [\hat j \hat k] [\hat i| + [\hat k \hat i] [\hat j| + [\hat i \hat j] [\hat k| \big)  |\eta]=0,
\end{equation}
where we defined the spinors $|\hat p]= p|1\rangle$.

The simplest example is the tree-level four-point MHV amplitude:
\begin{align}
{\mathcal A}_4^{(0)}(1^-,2^-,3^+,4^+) =i \frac{\langle 12 \rangle^3}{[\eta 1] \langle 13 \rangle \langle 14 \rangle} 
\bigg( & \frac{ [\eta 2] [34] f^{a_1 a_2 b} f^{b a_3 a_4}}{s_{12}} + \frac{[\eta 3] [42] f^{a_1 a_3 b} f^{b a_4 a_2}}{s_{13}} \nonumber \\ 
& + \frac{[\eta 4] [23] f^{a_1 a_4 b} f^{b a_2 a_3}}{s_{14}}\bigg),
\end{align}
where we chose particle $1$ to be the reference, as in \eqref{MHVlimit}.

%%%%%%%%%%%%%%%%%%%%%%%%%%%%%%%%%%%%%%%%
%%%%%%%%%%%%%%%%%%%%%%%%%%%%%%%%%%%%%%%%

\section{Manifestly dual integrands at one loop}
\label{sec:oneloop}

Now we will use the self-dual sector as a tool for understanding one-loop rational amplitudes in pure Yang-Mills theory. Let us start by describing the connection between the one-loop all-plus amplitudes and self-dual gauge theory. We will then show that the one-loop one-minus amplitudes are also related to self-dual gauge theory, in a manner analogous to MHV amplitudes at tree-level. We will also argue that the double copy relation to amplitudes in the so-called ${\mathcal N}=0$ supergravity is natural in these sectors. 

\subsection{Self-dual gauge theory and one-loop all-plus amplitudes}
\label{sec:counting}

\begin{figure}
\centering
\includegraphics[scale=0.4]{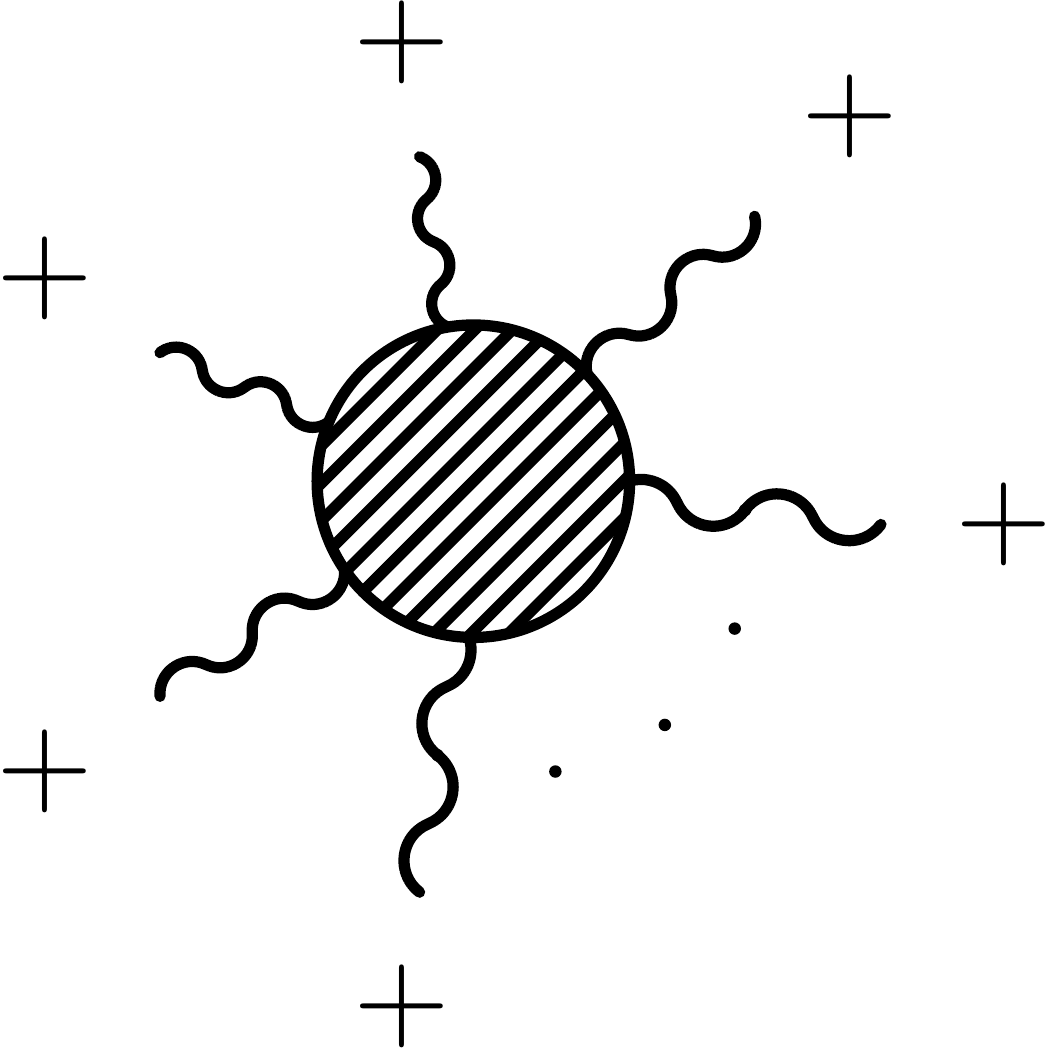}
\caption{An all-plus diagram.}
\label{fig:allPlus}
\end{figure}

It is well known that the all-plus one-loop amplitudes ${\mathcal A}^{(1)}(1^+, 2^+, \cdots, n^+)$ in Yang-Mills theory are computed by self-dual Yang-Mills theory~\cite{Cangemi:1996rx,Chalmers:1996rq}. It is instructive to see why this is. Recall the Yang-Mills Lagrangian in light-cone gauge presented in Eq.~\eqref{CSlag}. There are three kinds of vertex: the $(++-)$ and $(+--)$ three-point vertices, and the $(++--)$ four-point vertex. Now let us consider a one-loop all-plus diagram, as shown in Figure \ref{fig:allPlus}. 
We will show that such a one-loop amplitude contains only the $(++-)$ vertex, which is also the only vertex in the self-dual sector of the theory. To that end, let us suppose there are $n$ external gluons. Consider any Feynman diagram contributing to the loop amplitude, which contains $n_+$ vertices with helicities $(++-)$, $n_-$ vertices with helicities $(+--)$ and $n_4$ vertices with helicities $(++--)$. Let there be $I$ internal lines. As usual, the number of loops $L$ is related to the number of vertices and internal lines by
\begin{equation}
L = I - (n_+ + n_- + n_4) + 1 .
\end{equation}
For our application, we are interested in the case $L = 1$, so that the number of internal lines is simply the total number of vertices:
\begin{equation}
\label{counting1}
I = n_+ + n_- + n_4 . 
\end{equation}
Next, we count the number of plus and minus signs on the vertices. These lines must either be external lines, or must join to propagators. Therefore, we find
\begin{align}
n+I &= 2 n_+ + n_- + 2 n_4, \\
I &= n_+ + 2 n_- + 2n_4,
\label{counting3}
\end{align}
for plus and minus signs, respectively. From these equations it follows easily that
\begin{equation}
\label{eq:allPlusCount}
n_- + n_4 = 0,
\end{equation}
so that $n_- = 0$ and $n_4 = 0$. Thus, the one-loop amplitudes only include the $(++-)$ vertex, and can therefore be computed by restricting to the self-dual theory. 

We saw in the previous section what the Feynman rules in self-dual gauge theory are. There is a single type of vertex, $(++-)$, and the precise rule for that vertex was given in the first line of \eqref{LCvertices}. Using the Feynman rules, we can write numerators for any diagram. For instance, a four-point box diagram has the kinematic numerator\footnote{The factor of 2 comes from summing over the two helicity possibilities of the gluon running in the loop.}
\begin{equation}
\label{allplusboxnum}
n_{1|2|3|4} = \frac{2}{\langle \eta 1\rangle^2\langle \eta 2\rangle^2\langle \eta 3\rangle^2\langle \eta 4\rangle^2}
X(l,1)X(l+1,2)X(l+1+2,3)X(l-4,4),
\end{equation}
associated to the colour factor
\begin{equation}
\label{c1234}
c_{1|2|3|4} = f^{b_1 a_1 b_2} f^{b_2 a_2 b_3} f^{b_3 a_3 b_4} f^{b_4 a_4 b_1}.
\end{equation}
For any all-plus diagram, the factors which dress $X$ in the vertices -- see rule \eqref{LCvertices} -- combine with the polarisation vectors to produce an overall denominator $\Pi_{i=1}^n \langle \eta i\rangle^{-2}$. The interesting part of the numerators are the $X$ factors. We saw in the previous section that the factors $X$ satisfy Jacobi-like identities, which can be interpreted as Schouten identities. That statement is independent of whether the arguments of $X$ are on-shell or not. Therefore, the self-dual Feynman rules give colour-kinematics dual numerators for all-plus amplitudes in a straighforward manner. For instance,
\begin{eqnarray}
\label{jacobiboxtri}
n_{1|2|3|4} - n_ {2|1|3|4}\; && = \frac{2}{\langle \eta 1\rangle^2\langle \eta 2\rangle^2\langle \eta 3\rangle^2\langle \eta 4\rangle^2}
X(1,2)X(l,1+2)X(l+1+2,3)X(l-4,4) \nonumber \\
&& = n_{12|3|4},
\end{eqnarray}
where $n_{12|3|4}$ is the numerator of a triangle diagram with a massive corner, associated to the colour factor
\begin{eqnarray}
c_{1|2|3|4} - c_ {2|1|3|4} = f^{a_1 a_2 b_2} f^{b_1 b_2 b_3} f^{b_3 a_3 b_4} f^{b_4 a_4 b_1} = c_{12|3|4}.
\end{eqnarray}

In the same manner that boxes are associated to triangles, triangles are associated to bubbles through Jacobi identities. However, while bubbles are included in the representation of the integrand respecting the colour-kinematics duality, they vanish after integration. To see this, notice that the numerators of the bubbles depend on the loop momentum through
\begin{equation}
X(l,p)X(l+p,-p)=-X(l,p)^2=- (2p^X\cdot l)^2, \quad \textrm{where} \quad p^X_{\alpha \dot \alpha}=|\eta \rangle_\alpha (p|\eta \rangle)_{\dot \alpha}.
\end{equation}
The only tensor structures which can appear after the integration of $l^\mu l^\nu$ over the propagators are $g^{\mu\nu}$ and $p^\mu p^\nu$. Since $p^X$ is null and $p^X\cdot p=0$, it is clear that the integral vanishes. (A similar argument shows that the gravity bubbles obtained through the double copy formula also vanish.)

Let us also make a comment about dimensional regularisation. We are interested in having the loop momenta $L$ in $D=4-2\epsilon$ dimensions. However, the loop momentum $l$ in the factors $X(l,i)$ above is only the four-dimensional part of $L$. This procedure corresponds to the use of the four-dimensional helicity scheme \cite{Bern:1991aq}. The $(-2\epsilon)$-dimensional part of $L$ drops out from the numerator, but the propagators require the full $D$-dimensional loop momentum.  For instance, the propagator factors associated to the numerator \eqref{allplusboxnum} are
\begin{equation}
\frac{1}{D_{1|2|3|4}} = \frac{1}{L^2(L+p_1)^2(L+p_1+p_2)^2(L-p_4)^2}.
\end{equation}

The entire one-loop amplitude, for multiplicity $n$, is determined by the numerator of the $n$-gon integral, the other numerators being obtained from Jacobi identities. The $n$-gon numerator for the all-plus amplitude is
\begin{equation}
\label{allplusngon}
n_{1|2|3|\cdots|n} = 2\, (-1)^n 
\prod_{i=1}^n \frac{1}{\langle \eta i\rangle^2}\, X\Bigg(l+\sum_{j=1}^{i-1} i,i\Bigg).
\end{equation}

\subsection{One-loop one-minus amplitudes}

We have seen how to make the colour-kinematics duality manifest for all-plus amplitudes. Now we will see that there is a similar procedure for one-minus amplitudes. The relation between these two classes of amplitudes is analogous to the one between the self-dual sector and MHV amplitudes at tree level, reviewed earlier. Fortunately, the method used there will also be effective at one loop.

It is easy to modify the counting argument presented above for all-plus amplitudes so that it applies to one-minus amplitudes. If there are $n$ external particles, exactly one of which has negative helicity, the result is
\begin{equation}
n_- + n_4 =1.
\end{equation}
Therefore, if one can choose a gauge so that $n_- = 1$, the diagrams include only one vertex which is not of the self-dual type, and that vertex comes from the anti-self-dual sector, that is, a $(+--)$ vertex. This is exactly the same situation as for tree-level MHV amplitudes~\cite{Monteiro:2011pc}. As in that case, numerators satisfying Jacobi-like identities can be found by considering a reference particle with negative helicity. Let us say that the unique negative helicity particle is particle 1. Then we make the light-cone choice
\begin{equation}
|\eta\rangle \to |1\rangle.
\end{equation}
The resulting $n$-gon numerator, which determines the whole amplitude, is given by
\begin{equation}
\label{oneminusboxnum}
n_{1^-|2|3|\cdots|n} = 2\, (-1)^n \,\frac{1}{[\eta 1]^2} \, \overline X(l,1)
\prod_{i=2}^n \frac{1}{\langle \eta i\rangle^2}\, X\Bigg(l+\sum_{j=1}^{i-1} i,i\Bigg).
\end{equation}
The relation to the all-plus expression \eqref{allplusngon} is clear.
The associated Jacobi-like identities follow precisely in the same manner as Eqs.~\eqref{MHVjacobi} and \eqref{MHVshouten}.

We have thus succeeded in writing down rules to construct numerators satisfying the colour-kinematics duality for the one-minus amplitudes, for any number of external legs.

Let us remark that these numerators (both all-plus and one-minus) can be obtained using a scalar running in the loop, with the vertices \eqref{SGSvertices}. This is a consequence of the supersymmetric Ward identities for rational amplitudes \cite{Grisaru:1979re}.

\subsection{Double copy}

Having found a representation of one-loop gauge theory amplitudes satisfying the colour-kinematics duality, we can use the double copy formula \eqref{eq:double} to obtain one-loop gravity amplitudes. Two questions arise: we need to identify the specific theory of gravity resulting from the double copy; and we also need to verify that the procedure gives the correct amplitudes of that gravity theory.

On the first point, the gravity theory resulting from the ``squaring" of pure Yang-Mills theory is the so-called ${\mathcal N} = 0$ supergravity, consisting of Einstein gravity, a dilaton and a two-form field (which can be dualised into an axion in four dimensions). So the product of two scalar states (two helicities) in gauge theory gives four scalar states in gravity.

On the second point, consider the all-plus amplitudes first. All vertices in gauge theory that contribute to this amplitude are of the self-dual type, as we saw above. The double copy construction immediately yields a candidate gravity integrand for the helicity equal gravity amplitudes from this. In fact, precisely this expression follows directly from the self-dual gravity Lagrangian as derived from \cite{Siegel:1992wd}. Note that all the scalars appear minimally coupled in this action. Moreover, the purely gravity part of this action can be obtained by truncating the light-cone gravity action as written in e.g.  in Refs.~\cite{Ananth:2006fh,Ananth:2007zy,Ananth:2008ik} to fields which contain one field of one helicity type: only three point vertices in this gauge are of this type. This gravity Lagrangian can be extended to the supersymmetric Lagrangian presented \cite{Siegel:1992wd}. Truncating to $\mathcal{N}=0$ then shows that all scalars are minimally coupled.

The double copy also implies that, when going away from self-dual gravity to include one particle of opposite helicity, only one other type of vertex is needed. This is supported by a counting argument for vertices, which closely follows the argument detailed above for gauge theory. In fact, it is precisely the same argument, with the difference that there is an infinite sequence of vertices in the gravity light-cone Lagrangian \cite{Ananth:2006fh,Ananth:2007zy,Ananth:2008ik}. Instead of $n_4$, we consider $n_{\sigma_+\sigma_-}$, which is the number of vertices with more than three-points involving $\sigma_+$ plus-helicity particles and $\sigma_-$ minus-helicity particles; such vertices always possess at least two particles of each helicity, so $\sigma_\pm \geq 2$. Then, the counterparts of \eqref{counting1}-\eqref{counting3} are
\begin{align}
I &= n_+ + n_- + \sum_{\sigma_+,\sigma_-}  n_{\sigma_+\sigma_-}, \\
n-\varepsilon +I &= 2 n_+ + n_- + \sum_{\sigma_+,\sigma_-} \sigma_+ \,n_{\sigma_+\sigma_-}, \\
\varepsilon + I &= n_+ + 2 n_- + \sum_{\sigma_+,\sigma_-} \sigma_- \,n_{\sigma_+\sigma_-},
\end{align}
where $\varepsilon=0$ for all-plus diagrams and $\varepsilon=1$ for one-minus diagrams. We conclude that
\begin{equation}
n_- + \sum_{\sigma_+,\sigma_-} (\sigma_- -1) \,n_{\sigma_+\sigma_-} =\varepsilon.
\end{equation}
In the case of all-plus amplitudes, this argument shows that there can be no other vertices than the self-dual type, just as the double copy yields. In the case of one-minus amplitudes, there must be one more vertex, either of anti-self-dual type ($n_-=1$) or a higher-point vertex with two negative helicity particles ($n_{\sigma_+2}=1$). The double copy indicates that we can choose $n_-=1$ by specifying a gauge as in gauge theory: $|\eta\rangle \to |1\rangle$, where 1 is the single negative helicity particle. For this to be possible, it is required that no higher-point gravity vertex of the type $(\sigma_+,2)$ leads to poles in $\langle \eta 1 \rangle$. Under this assumption, which is true through at least five points \cite{Ananth:2008ik}, the gravity counting argument agrees with the double copy.

Note that the difference between self-dual $\mathcal{N}=0$ and pure self-dual gravity are minor. For the rational amplitudes they only differ by a factor of $2$: Let us recall that the supersymmetric Ward identities imply that we can replace gravitons and other massless particles running in the loop with massless scalars \cite{Grisaru:1979re}, see also \cite{Bern:1993wt}. As a consequence, we have that
\begin{equation}\label{BernDegFreedom}
{\mathcal M}_n^{(1),\textrm{any states}}(1^\pm,2^+,\ldots,n^+) = N_s\, {\mathcal M}_n^{(1),\textrm{one scalar}}(1^\pm,2^+,\ldots,n^+) ,
\end{equation}
where $N_s$ is the number of states running in the loop (bosonic minus fermionic). For pure gravity, there are two helicities, so $N_s=2$. For ${\mathcal N} = 0$ supergravity, we have $N_s=4$. In the self-dual sector of gravity it can be seen from \cite{Siegel:1992wd} that the scalars in $\mathcal{N}=0$ SUGRA are minimally coupled. Comparing the explicit expression for the one-loop integrand with a scalar running to a loop with a graviton running shows the two expressions are equivalent. Hence in this theory there is a $N_s$ factor diagram-by-diagram.

In conclusion, we have presented a Lagrangian based proof that in the helicity-equal sector at one loop the double copy conjecture is true. In the one-helicity unequal sector at one loop this was proven to five points, while it is plausible it holds to all multiplicity.

\subsection{Examples}
\label{subsec:examples}

To illustrate the general arguments we presented above, we will now explicitly describe some simple examples. In particular, we will consider four-point gravity amplitudes to verify the double copy procedure. As a first step, we will calculate three-point one-leg-off-shell currents in gauge theory and gravity.

The higher-point amplitudes obtained in the same way will obviously satisfy kinematic consistency conditions, i.e. as they are rational they will not have any four-dimensional cuts and applying (a number of) collinear limits they will reduce to the four-point result.

\subsection*{$(+++)$ one-loop current}

A useful warmup for the (++++) one-loop gravity example is to calculate the $(+++)$ one-loop Yang-Mills current, and then the corresponding one-loop graviton current.
The latter will be calculated from the YM expression by the double copy construction, using $(++-)$ vertices as a building block. We will closely follow the procedure explained by Brandhuber, Spence and Travaglini in \cite{Brandhuber:2006bf} to calculate the (++++) one-loop amplitude in non-supersymmetric YM. 

\begin{figure}[h!]
\centering
\includegraphics[scale=0.4]{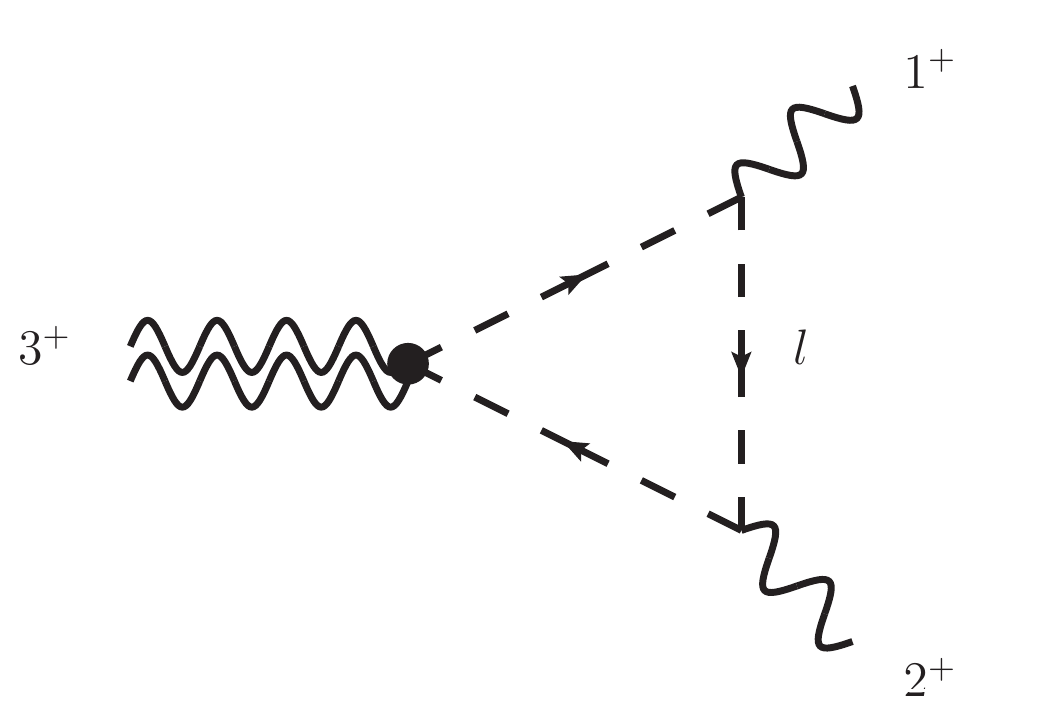}
\caption{The one-loop $(+++)$ current of Yang-Mills theory. Leg 3 is off-shell.}
\label{fig:3plusvertex}
\end{figure}

The $(+++)$ one-loop YM current is depicted in Figure~\ref{fig:3plusvertex}. We have already shown that bubble diagrams do not contribute. Using the $(++-)$-vertex and the polarization factors for particles 1 and 2, \eqref{LCvertices}-\eqref{LChel}, the diagram is written as
\begin{equation}
 J_3^{(1)}=\int \frac{d^DL}{(2\pi)^{D}}\frac{\langle\eta|l_1|1]\langle\eta|l_2|2]X(l_3,3)}{\langle\eta1\rangle\langle\eta2\rangle 3_\eta}\cdot\frac{1}{L_1^2L_2^2L_3^2},
\label{ym3plus}
\end{equation}
with $L_1^2=L^2$, $L_2^2=(L-p_2)^2$, and $L_3^2=(L+p_1)^2$. The integral is evaluated in $D=4-2\epsilon$ dimensions. As noted above, the scheme employed here is such that the vertices live in four dimensions, whereas the propagators live in $D$ dimensions. The loop momentum $L$ is decomposed into (orthogonal) four- and $(-2\epsilon)$-dimensional parts as 
\begin{equation}
 L=l+l_{-2\epsilon} \qquad \text{with} \qquad  L^2=l^2+l^2_{-2\epsilon}=l^2-\mu^2,
\label{Ddimloopmom}
\end{equation}
where the $(-2\epsilon)$-dimensional part corresponds to a mass $\mu^2$ of a scalar running in the loop. The quantity
\begin{equation}
 \frac{\langle\eta|l_1|1]\langle\eta|l_2|2]}{\langle\eta1\rangle\langle\eta2\rangle}
\end{equation}
appearing in the integrand can be simplified further, and the dependence on the $(-2\epsilon)$-dimensional subspace can be extracted. To do so, write it as
\begin{align}
 \frac{\langle\eta|l_112l_1|\eta\rangle}{\langle\eta1\rangle\langle\eta2\rangle\langle12\rangle} &=-l_1^2\frac{[12]}{\langle12\rangle}-\frac{2(l_1\cdot p_1)\langle\eta|1\,l_1|\eta\rangle}{\langle\eta1\rangle\langle\eta2\rangle\langle12\rangle}-\frac{2(l_1\cdot p_2)\langle\eta|l_1\,2|\eta\rangle}{\langle\eta1\rangle\langle\eta2\rangle\langle12\rangle},
\end{align}
where the definition of the Clifford algebra has been used on $\slashed{l_1}\,\slashed{1}$ and $\slashed{2}\,\slashed{l_1}$ on the left hand side. The scalar products can be expressed as the difference of propagators
\begin{equation}
 2(l_1\cdot p_1)=l_3^2-l^2_1=L_3^2-L_1^2 \quad \text{and} \quad 2(l_1\cdot p_2)=L_1^2-L_2^2,
\end{equation}
and $l_1^2=L_1^2+\mu^2$ by \eqref{Ddimloopmom}. Finally, one obtains
\begin{equation}
\label{extractmu}
 \frac{\langle\eta|l_1|1]\langle\eta|l_2|2]}{\langle\eta1\rangle\langle\eta2\rangle}=-\mu^2\frac{[12]}{\langle12\rangle}+\frac{Q}{\langle12\rangle}
\;\; ; \;\;
 Q=L_1^2\frac{\langle\eta|l_3(1+2)|\eta\rangle}{\langle\eta1\rangle\langle\eta2\rangle}+L_2^2\frac{\langle\eta|l_1|1]}{\langle\eta2\rangle}+L_3^2\frac{\langle\eta|l_1|2]}{\langle\eta1\rangle}.
\end{equation}
The one-loop current becomes
\begin{equation}
 J_3^{(1)}(+++)=\int \frac{d^DL}{(2\pi)^{D}}\Big(-\mu^2\frac{[12]}{\langle12\rangle}+\frac{Q}{\langle12\rangle}\Big)\frac{X(l_3,3)}{3_\eta}\cdot\frac{1}{L_1^2L_2^2L_3^2}.
\end{equation}
Upon standard one-loop integration of this expression, one finds that only the part proportional to $\mu^2$ survives, and that it can be related to an integral in $D+2$ dimensions \cite{Bern:1995db}. Eventually, the result is
\begin{equation}
 J_3^{(1)}(+++)=\frac{i}{(4\pi)^{2-\epsilon}}\cdot\frac{\Gamma(1+\epsilon)\Gamma(1-\epsilon)^2}{\Gamma(4-2\epsilon)(-p^2_3)^\epsilon}\cdot\frac{[12]^2}{\langle12\rangle}\cdot\frac{\langle\eta 1\rangle \langle\eta 2\rangle}{3_\eta}.
\label{ymvert}
\end{equation}

Next consider gravity. We have shown before that the colour-kinematics duality is manifest in gauge theory, so we can just apply the double copy by squaring the numerator (that is, the integrand excluding the propagators) in equation \eqref{ym3plus}. Again, one does not need to consider bubble topologies, which vanish upon integration. So the three-graviton all-plus one-loop current is given by
\begin{equation}
 \mathcal{J}_3^{(1)}(+++)=\int \frac{d^DL}{(2\pi)^{D}}\Big(\frac{\langle\eta|l_1|1]\langle\eta|l_2|2]X(l_3,3)}{\langle\eta1\rangle\langle\eta2\rangle3_\eta}\Big)^2\cdot\frac{1}{L_1^2L_2^2L_3^2},
\label{grav3plus}
\end{equation}
which will become, using \eqref{extractmu},
\begin{equation}
  \mathcal{J}_3^{(1)}(+++)=\int \frac{d^DL}{(2\pi)^{D}}\Big(-\mu^2\frac{[12]}{\langle12\rangle}+\frac{Q}{\langle12\rangle}\Big)^2\frac{X(l_3,3)^2}{3_\eta^2}\cdot\frac{1}{L_1^2L_2^2L_3^2}.
\end{equation}
Expanding out the integrand gives terms proportional to $\mu^4$, $\mu^2$, and $\mu^0$. It can be shown easily that only the $\mu^4$ term survives after integration, i.e. the integral simplifies to
\begin{equation}
\mathcal{J}_3^{(1)}(+++)=\int \frac{d^DL}{(2\pi)^{D}}\mu^4\frac{[12]^2}{\langle12\rangle^2}
\frac{X(l_3,3)^2}{3_\eta^2}\cdot\frac{1}{L_1^2L_2^2L_3^2}.
\end{equation}
Rewriting this integral as a higher-dimensional integral gives
\begin{equation}
\begin{split}
\label{gravvert}
 \mathcal{J}_3^{(1)}(+++)&=(-\epsilon)(1-\epsilon)(4\pi)^2\int \frac{d^{D+4}L}{(2\pi)^{D+4}}\frac{[12]^2}{\langle12\rangle^2}\frac{X(l_3,3)^2}{3_\eta^2}\cdot\frac{1}{L_1^2L_2^2L_3^2}\\&=\frac{i}{(4\pi)^{2-\epsilon}}\cdot\frac{2\Gamma(1+\epsilon)\Gamma(2-\epsilon)^2}{\Gamma(7-2\epsilon)(-p_3^2)^{-1+\epsilon}}\cdot \left(\frac{[12]^2}{\langle12\rangle}\cdot\frac{\langle\eta 1\rangle \langle\eta 2\rangle}{3_\eta}\right)^2.
\end{split}
\end{equation}
Up to a factor of $(p_3^2)^{-1}$ and numerical coefficients, the gravity current turns out to be the square of the Yang-Mills current \eqref{ymvert}. Note that we have not cared for possible different internal helicity configurations in the example above since it was meant to sketch the general procedure of how the double copy construction works. We will comment on how to properly take internal helicities into account in the next section when we also consider box topologies. Using the arguments to be presented below it will become clear that the gauge theory current actually has to be multiplied by a factor of 2 and the gravity one by a factor of 4.
 
\subsection*{(++++) one-loop amplitude}
In this subsection, the four-point all-plus one-loop gravity amplitude ${\mathcal M}_4^{(1)}$ will be calculated using the BCJ double copy construction. To do so, we first write the corresponding full colour-dressed YM amplitude in BCJ form, which is pictorially given by
\begin{equation}
\label{bcjym}
A_{4}^{(1)}(+++\, +)=\int\frac{d^DL}{(2\pi)^D}\vcenter{\hbox{\includegraphics[width=0.6\textwidth]{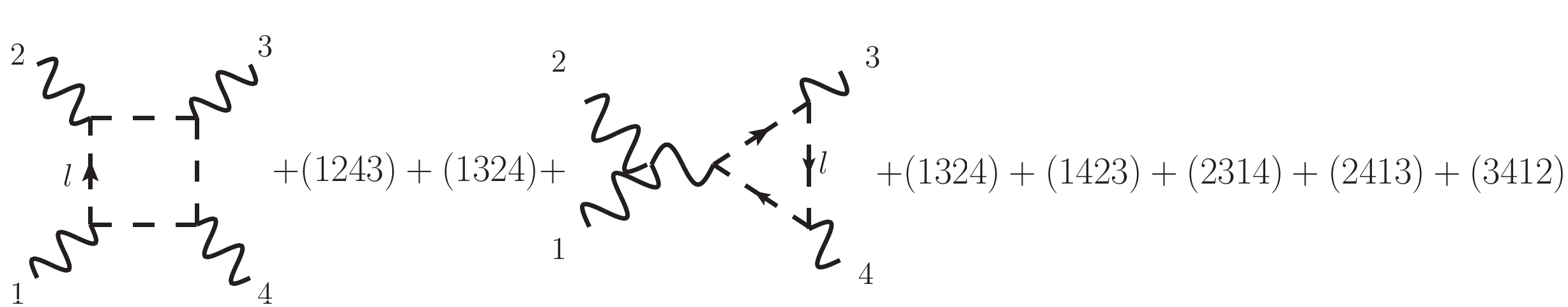}}}
\end{equation}
Here again bubble integrals have been ignored, as they will integrate to zero in the YM case and also after squaring. Note that one has to be aware of a subtlety here. There are two possible internal helicity configurations for the integrand of each topology. This corresponds to the BCJ numerators corresponding to a sum of two terms that each individually are dual, i.e. for the boxes one roughly has \begin{equation}
\text{Box}\sim f^{box}\frac{n^{box}}{D(l_{box})}=f^{box}\frac{(n^B_a+n^B_b)}{D(l_{box})}
\end{equation}
where $n_a$ and $n_b$ are the numerators for the two possible internal configurations. For the triangles one has a relative minus sign because of Bose symmetry, i.e.
\begin{equation}
\text{Tri}=f^{tri}\frac{n^{tri}}{D(l_{tri})}=f^{tri}\frac{(n^T_a-n^T_b)}{D(l_{tri})}.
\end{equation}
Consequently in the double copy construction one has to square these numerators. But since they are related by
\begin{equation}\label{factorfour}
n^B_a = n^B_b \quad n^T_a=-n^T_b,
\end{equation}
as can be easily shown using the properties of the $X$ vertices, one finds after squaring
\begin{equation}
\text{boxes}+\text{triangles} = 4\Big(\frac{(n^B_a)^2}{D(l_{box})}+\frac{(n^T_a)^2}{D(l_{tri})}\Big).
\end{equation}
This simply means that one can do the doubly copy construction considering only one internal helicity configuration and multiply the result by a factor of four. This is essentially equation \eqref{BernDegFreedom}, i.e. this factor corresponds to the bosonic degrees of freedom running in the loop of extended supergravity. As discussed above they are the two graviton states, a dilation, and an antisymmetric two-form.
As the numerators satisfy kinematic Jacobi relations by construction, they can immediately be squared, i.e. the $(++-)$ vertices will be squared. Having done this consider, consider first the box terms, e.g. for example the ordering $1234$.  This part of the gravity amplitude is
\begin{equation}
\text{Box}(1234)=\vcenter{\hbox{\includegraphics[scale=0.5]{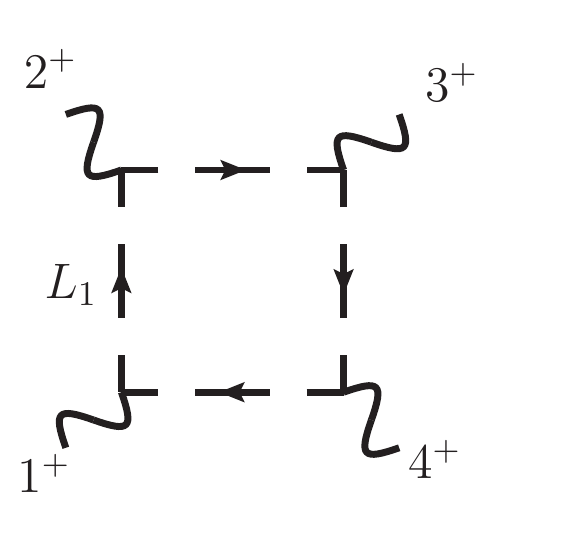}}}\!\!\!\!\!\!\!\!\!\!\!\!=\int\frac{d^DL}{(2\pi)^{D}}\Big(\frac{\langle\eta|l_1|1]}{\langle\eta1\rangle}\frac{\langle\eta|l_2|2]}{\langle\eta2\rangle}\frac{\langle\eta|l_3|3]}{\langle\eta3\rangle}\frac{\langle\eta|l_4|4]}{\langle\eta4\rangle}\Big)^2\cdot\frac{1}{L_1^2L_2^2L_3^2L_4^2},
\end{equation}
with $L_1^2=L^2$, $L_2^2=(L-p_2)^2$, $L_3^2=(L-p_2-p_3)^2$, and $L_4^2=(L+p_1)^2$. Similarly to \eqref{extractmu}, one can rewrite the terms of the integrand as
\begin{equation}\label{allplusboxsimple}
\begin{split}
&\frac{\langle\eta|l_1|1]\langle\eta|l_2|2]}{\langle\eta1\rangle\langle\eta2\rangle}=-\mu^2\frac{[12]}{\langle12\rangle}+\frac{Q}{\langle12\rangle}
\;\; ; \;\;
 Q=L_1^2\frac{\langle\eta|l_3(1+2)|\eta\rangle}{\langle\eta1\rangle\langle\eta2\rangle}+L_2^2\frac{\langle\eta|l_1|1]}{\langle\eta2\rangle}+L_3^2\frac{\langle\eta|l_1|2]}{\langle\eta1\rangle},
\\&
\frac{\langle\eta|l_3|3]\langle\eta|l_4|4]}{\langle\eta3\rangle\langle\eta4\rangle}=-\mu^2\frac{[34]}{\langle34\rangle}+\frac{\tilde Q}{\langle34\rangle}
\;\; ; \;\;
 \tilde Q=L_3^2\frac{\langle\eta|l_2(3+4)|\eta\rangle}{\langle\eta3\rangle\langle\eta4\rangle}+L_2^2\frac{\langle\eta|l_3|4]}{\langle\eta3\rangle}+L_4^2\frac{\langle\eta|l_3|3]}{\langle\eta4\rangle},
\end{split}
\end{equation}
so that the integral becomes
\begin{equation}
 \text{Box}(1234)=\int\frac{d^DL}{(2\pi)^{D}}\Big(-\mu^2\frac{[12]}{\langle12\rangle}+\frac{Q}{\langle12\rangle}\Big)^2\Big(-\mu^2\frac{[34]}{\langle34\rangle}+\frac{\tilde Q}{\langle34\rangle}\Big)^2\cdot\frac{1}{L_1^2L_2^2L_3^2L_4^2}.
\label{boxygrav}
\end{equation}
Focus on the $\mu^8$-part of this expression. It will now be shown that this piece is proportional to the gravity result. The $\mu^8$-part is given by
\begin{equation}
 \text{Box}(1234)|_{\mu^8}=\int\frac{d^DL}{(2\pi)^{D}}\mu^8\frac{[12]^2[34]^2}{\langle12\rangle^2\langle34\rangle^2}\cdot\frac{1}{L_1^2L_2^2L_3^2L_4^2}=\frac{[12]^2[34]^2}{\langle12\rangle^2\langle34\rangle^2}I^{1234}_{D=4-2\epsilon}[\mu^8]
\end{equation}
where $I^{1234}_{D=4-2\epsilon}[\mu^8]=(-\epsilon)(1-\epsilon)(2-\epsilon)(3-\epsilon)(4\pi)^4 I^{1234}_{D=12-2\epsilon}[1]$ is a scalar integral in $D=12-2\epsilon$ dimensions \cite{Bern:1995db}. Similar computations can be done for the other two box configurations, and one finds
\begin{equation}
 \text{Box}|_{\mu^8}=\frac{[12]^2[34]^2}{\langle12\rangle^2\langle34\rangle^2}\Big(I^{1234}_{D=4-2\epsilon}[\mu^8]+I^{1243}_{D=4-2\epsilon}[\mu^8]+I^{1324}_{D=4-2\epsilon}[\mu^8]\Big),
\end{equation}
which is the one-loop four-point all-plus gravity amplitude as computed by Bern et al \cite{Bern:1998sv} up to a factor of four. However, the factor of four follows from the discussion at the beginning of this section \eqref{factorfour} so that the $\mu^8$ piece does actually give the correct all-plus gravity amplitude for $\mathcal{N}=0$ supergravity.

In order to check that our computation gives the correct result, it must be verified that the $\mu^6$, $\mu^4$, $\mu^2$, and $\mu^0$ terms of the box integral \eqref{boxygrav} and the triangle integrals cancel, i.e
\begin{equation}\label{vanishingsum}
 \text{Box}|_{\mu^6}+\text{Box}|_{\mu^4}+\text{Box}|_{\mu^2}+\text{Box}|_{\mu^0}+\text{Triangles}\stackrel{?}{=}0
\end{equation}
The box terms can be extracted from \eqref{boxygrav}. The triangle diagram contributions are obtained by putting together a $(++-)$ tree-level current and the one-loop current \eqref{gravvert}. After a bit of algebra, one finds
 \begin{equation}
   \text{Tri}(1234)=\vcenter{\hbox{\includegraphics[scale=0.4]{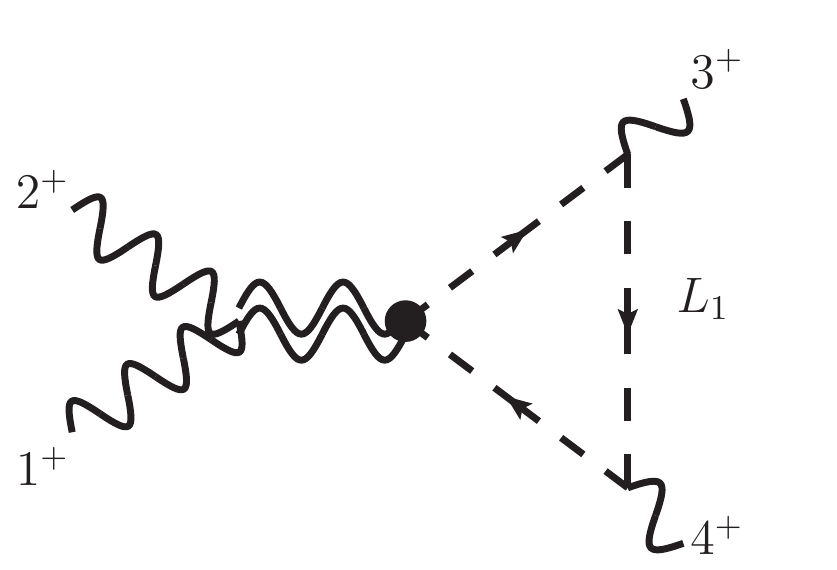}}}\!\!\!\!\!=\frac{i}{(4\pi)^{2-e}}\frac{2\Gamma(1+\epsilon)\Gamma(2-\epsilon)^2}{\Gamma(7-2\epsilon)(-s_{12})^{\epsilon}}\frac{\langle\eta|34|\eta\rangle^4}{\langle12\rangle^2\langle34\rangle^2\prod_{i=1}^4\langle\eta i\rangle^2}.
 \end{equation}
The other triangle configurations are obtained by permutation of the external legs. Finally, after evaluating all these integrals, we checked numerically (up to and including $\mathcal{O}(\epsilon^2)$ in dimensional regularization) that the terms in \eqref{vanishingsum} indeed add up to zero. 

In summary, we have calculated the one-loop (++++) $\mathcal{N}=0$ supergravity amplitude ${\mathcal M}_4^{(1)}$ using the BCJ double copy construction, and reproduced the well-known expression (taking into account the factor of four discussed above)
\begin{equation}
\begin{split}
\label{M4allplus}
{\mathcal M}_4^{(1)}(+++\,+)&=4\;\text{Box}|_{\mu^8}=4\frac{[12]^2[34]^2}{\langle12\rangle^2\langle34\rangle^2}\Big(I^{1234}_{D=4-2\epsilon}[\mu^8]+I^{1243}_{D=4-2\epsilon}[\mu^8]+I^{1324}_{D=4-2\epsilon}[\mu^8]\Big).
\end{split}
\end{equation}

\subsection*{$(-++)$ one-loop current}
As another interesting example of numerators satisfying the colour-kinematics duality, the one-minus one-loop gravity three-current and four-point amplitude will be calculated. To do so, one can reuse most parts of the machinery from the previous computation. To make the colour-kinematics duality manifest, we implement the gauge choice introduced earlier
\begin{equation}
| \eta \rangle = | 1 \rangle,
\end{equation}
where particle 1 has negative helicity particle. This choice eliminates four-point vertices and forces particle $1$ to couple to a $(--+)$ vertex.

\begin{figure}[h!]
\centering
\includegraphics[scale=0.4]{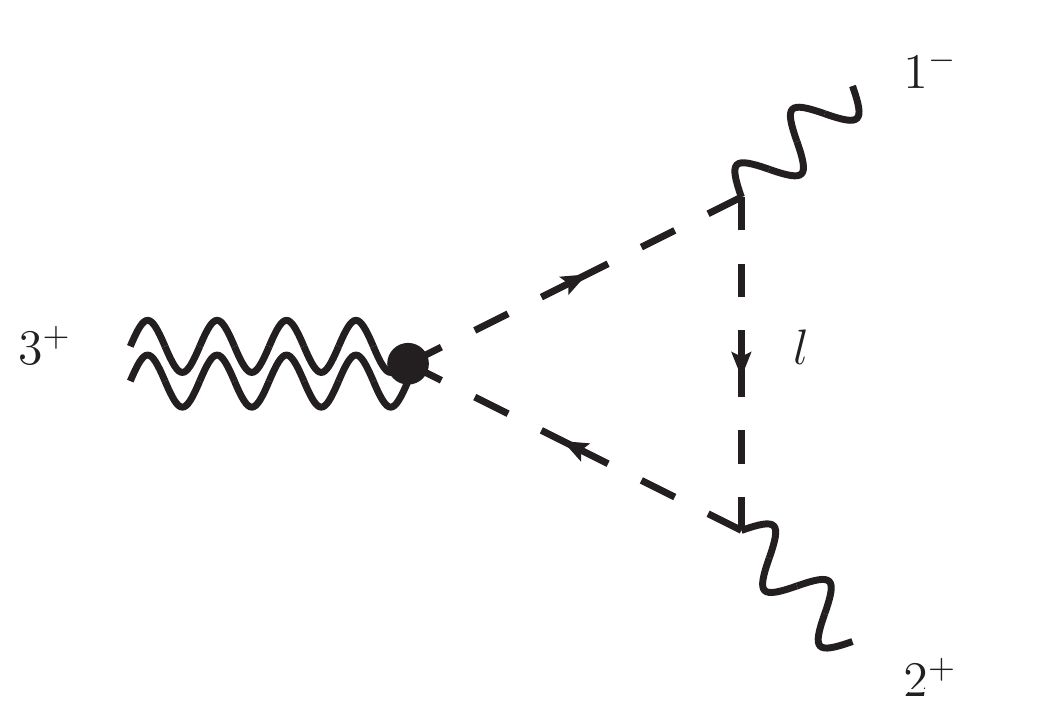}
\caption{The one-loop $(-++)$ current of Yang-Mills theory. Leg 3 is off-shell.
}
\label{fig:3oneminusvertex}
\end{figure}
We begin with the $(-++)$ one-loop current as it is a building block for the one minus four-point amplitude. Using the rules \eqref{LCvertices}-\eqref{LChel}, including the polarization factors for the on-shell particles 1 and 2, and taking into account the choice for $|\eta \rangle$, one arrives at
\begin{equation}
{J}_3^{(1)}(-++)=\int \frac{d^DL}{(2\pi)^{D}}\Big(\frac{[\eta|l_1|1\rangle \langle 1|l_2|2]X(l_3,3)}{[\eta 1]\langle 1 2\rangle 3_\eta}\Big)\cdot\frac{1}{L_1^2L_2^2L_3^2} .
\label{gauge3oneminus}
\end{equation}
The numerator depends on the loop momenta through
\begin{eqnarray}
[\eta|l_1|1\rangle \langle 1|l_2|2]X(l_3,3)=4 \zeta_{1\mu}\zeta_{2\nu}\zeta_{3\sigma} \, l_1^\mu  l_2^\nu  l_3^\sigma
= 4 \langle 12\rangle \zeta_{1\mu}\zeta_{2\nu}\zeta_{2\sigma} \, l_1^\mu  l_1^\nu  l_1^\sigma,
\end{eqnarray}
where $\zeta_1=|1\rangle [\eta|=\eta$, $\zeta_2=|1\rangle [2|$ and $\zeta_3=|1\rangle (\langle 1|3)=\langle 12\rangle \zeta_2$.
The tensorial structures appearing after the integration of $ l_1^\mu  l_1^\nu  l_1^\sigma$ over the propagators can only be of six types: $g^{\mu\nu}p_1^\sigma$, $g^{\mu\nu}p_2^\sigma$, $ p_1^\mu  p_1^\nu  p_1^\sigma$, $ p_1^\mu  p_1^\nu  p_2^\sigma$, $ p_1^\mu  p_2^\nu  p_2^\sigma$ and $ p_2^\mu  p_2^\nu  p_2^\sigma$ (recall that $p_3=-p_1-p_2$). Now, the vectors $\zeta_i$ are null and mutually orthogonal. Moreover, $\zeta_i\cdot p_1=0$ and $\zeta_2\cdot p_2=0$. Therefore, there is no other possibility than
\begin{equation}
{J}_3^{(1)}(-++)=0.
\end{equation}
So this current does not play a role in gauge theory amplitudes.

Similarly, for the corresponding gravity current obtained through the double copy formula, we get
\begin{equation}
\mathcal{J}_3^{(1)}(-++)=0.
\end{equation}

\subsection*{$(-+++)$ one-loop amplitude}

In this subsection, the four-point one-minus one-loop gravity amplitude ${\mathcal M}_4^{(1)}$ will be calculated using the BCJ double copy construction. Again, we first write the one-minus one-loop YM amplitude in a BCJ form which is pictorially given by the same expansion as in the all-plus case \eqref{bcjym}, except for diagrams where particle $1^-$ is attached to the corner of a triangle. We checked above that the latter diagrams vanish after integration. Additionally, bubbles integrate to zero so will be ignored as before.
The gravity box diagram is given by squaring the corresponding gauge theory numerators, namely
\begin{equation}
\text{Box}(-+++)=\int\frac{d^DL}{(2\pi)^{D}}\left(\frac{[\eta|l_1|1\rangle}{[\eta1]}\frac{\langle1|l_2|2]}{\langle12\rangle}\frac{\langle1|l_3|3]}{\langle13\rangle}\frac{\langle1|l_4|4]}{\langle14\rangle}\right)^2\cdot\frac{1}{L_1^2L_2^2L_3^2L_4^2}.
\end{equation}
The first two factors can be rewritten using $[\eta|l_1|1\rangle=2l_1\cdot\eta$ and $l_2|2]=l_1|2]$, whereas the last two can be rewritten as in the all-plus case, using \eqref{extractmu}. One finds
\begin{equation}
\text{Box}(-+++)=\int\frac{d^DL}{(2\pi)^{D}}\left(\left(\frac{2l_1\cdot\eta\langle1|l_1|\eta]}{[\eta1]\langle12\rangle}\Big)\Big(-\mu^2\frac{[34]}{\langle34\rangle}+\frac{\tilde Q}{\langle34\rangle}\right)\right)^2\cdot\frac{1}{L_1^2L_2^2L_3^2L_4^2},
\end{equation}
with $\tilde Q$ defined in \eqref{allplusboxsimple}. Note that the highest power in $\mu^2$ appearing for this helicity configuration is two, in contrast to four in the all-plus case.

The triangle diagrams contributing can be constructed from \eqref{gravvert} by multiplying it with the appropriate gravity $(--+)$ vertex, i.e. the square of the second vertex in \eqref{LCvertices}. One finds 
 \begin{equation}
   \text{Tri}(-+++)=\vcenter{\hbox{\includegraphics[scale=0.4]{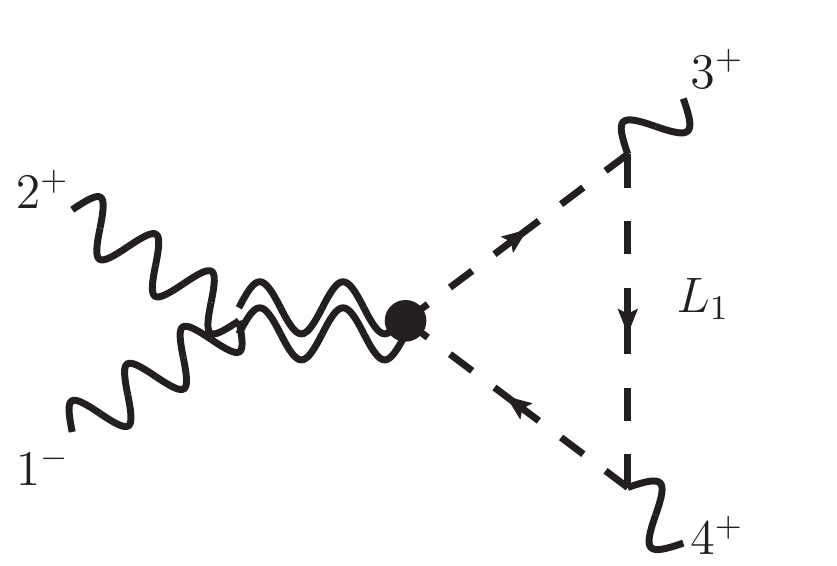}}}\!\!\!\!\!=\frac{i}{(4\pi)^{2-\epsilon}}\frac{2\Gamma(1+\epsilon) \Gamma(2-\epsilon)^2}{\Gamma(7-2\epsilon)\left(-s_{12}\right){}^\epsilon} \frac{ \langle 1|34|1\rangle ^4 [2\eta ]^2}{ \langle 34\rangle ^2 \langle 12\rangle^2\langle13\rangle^2\langle14\rangle^2. [1\eta ]^2}
 \end{equation}

Adding up the contributions from boxes and triangles, evaluating the integrals numerically,  and finally taking into account \eqref{factorfour} we find nice agreement with the literature result \cite{Bern:1993wt}
\begin{equation}
{\mathcal M}_4^{(1)}(-+++)=4\Big(\frac{st}{u}\Big)^2\Big(\frac{[24]^2}{[12]\langle23\rangle\langle34\rangle[41]}\Big)^2\Big(\frac{s^2+st+t^2}{5760}\Big)
\end{equation}
in the limit $\epsilon\rightarrow 0$. In other words, we have also constructed the $(-+++)$ one-loop $\mathcal{N}=0$ gravity amplitude using the double copy formula.

%%%%%%%%%%%%%%%%%%%%%%%%%%%%%%%%%%%%%
%%%%%%%%%%%%%%%%%%%%%%%%%%%%%%%%%%%%%

\section{Colour-kinematics duality after integration}
\label{sec:integrated}

We have seen that the self-dual kinematic algebra leads to natural BCJ numerators for two families of one-loop amplitudes in pure Yang-Mills theory: the all-plus amplitudes and one-minus amplitudes. These amplitudes are special because they have vanishing four-dimensional cuts, and are therefore purely rational. Badger \cite{Badger:2008cm} presented a method to evaluate rational terms using $D$-dimensional cuts. Using this procedure, we will show that the appearance of the kinematic algebra at the level of the {\it integrated} amplitude explains certain linear relations among partial amplitudes found by Bjerrum-Bohr et al \cite{BjerrumBohr:2011xe}. These relations have been proven in \cite{Boels:2011tp}. The goal here is to look for residual algebraic structure from the colour-kinematics duality after integration.  

It was shown in \cite{BjerrumBohr:2011xe,Boels:2011tp} that one-loop all-plus partial amplitudes satisfy a set of linear relations which resemble the tree-level Kleiss Kuijf relations. An example, at five points, is
\begin{align}
0 ={}& A_{5;1}^{(1)}( 1, 4, 3, 5,2) + A_{5;1}^{(1)}(1, 5, 3, 4,2) +
 A_{5;1}^{(1)}(1,2, 3, 4, 5) \nonumber \\
&+ A_{5;1}^{(1)}(1,2, 3, 5, 4) +
 A_{5;1}^{(1)}(1, 5, 3,2,4) + A_{5;1}^{(1)}(1, 4, 3,2,5)\,,
\label{eq:KKlike}
\end{align}
where the subscript 1 denotes that these partial amplitudes are planar (correspond to a single colour trace).
It was noted that these relations could be explained by a structure of ``vertices" with certain symmetry properties. At $n$-points, each diagram contributing to the amplitude would possess a single completely-symmetric four-point ``vertex" ${\mathcal D}^{q_1q_2q_3q_4}$, and $n-4$ completely-antisymmetric three-point ``vertices" ${\mathcal F}^{q_1q_2q_3}$. For instance, at five points,\footnote{The propagators and the sign of the momentum connecting ${\mathcal F}$ and ${\mathcal D}$ were not included in \cite{BjerrumBohr:2011xe}. Their introduction is natural, however, if we look for a representation of the ``vertices".}
\begin{eqnarray}
\label{eq:5ptsFD}
A_{5;1}^{(1)}(1,2,3,4,5) &\longrightarrow& {\mathcal F}^{k_1k_2(-q)} \frac{i}{s_{12}}{\mathcal D}^{qk_3k_4k_5} + {\mathcal F}^{k_5k_1(-q)}\frac{i}{s_{51}} {\mathcal D}^{qk_2k_3k_4} +{\mathcal F}^{k_4k_5(-q)}\frac{i}{s_{45}} {\mathcal D}^{qk_1k_2k_3} \nonumber \\
&&+{\mathcal F}^{k_3k_4(-q)}\frac{i}{s_{34}} {\mathcal D}^{qk_5k_1k_2} +{\mathcal F}^{k_2k_3(-q)}\frac{i}{s_{23}} {\mathcal D}^{qk_4k_5k_1} \,.
\end{eqnarray}
Identity \eqref{eq:KKlike} follows directly from this symmetry or antisymmetry of the vertices. The same happens for higher $n$, where one would always have four ``currents" made from ${\mathcal F}$'s (and propagators) meeting at a ``vertex" ${\mathcal D}$.

\begin{figure}[h!]
\centering
\includegraphics[scale=0.5]{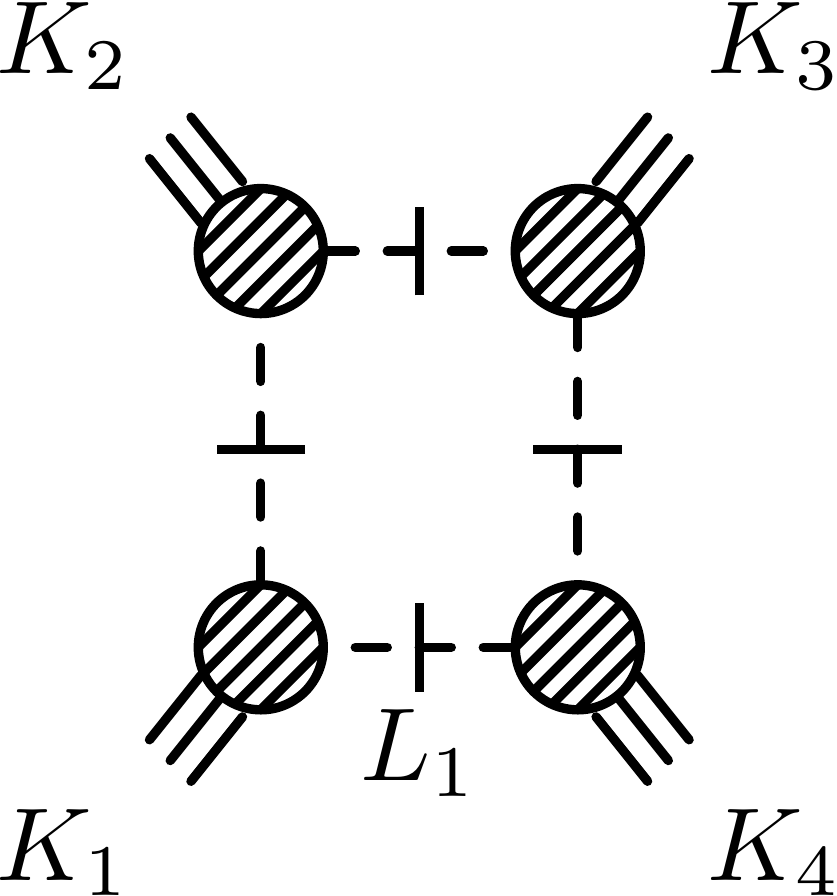}
\caption{A box cut of an $n$ point amplitude.}
\label{fig:cutbox}
\end{figure}

Let us see how this structure follows naturally from the kinematic algebra, and from the fact that we are considering rational amplitudes, that is, amplitudes with vanishing four-dimensional cuts. Using the method of $D$-dimensional cuts of \cite{Badger:2008cm}, an all-plus amplitude is given by a sum over cut boxes. Consider the box represented in Figure~\ref{fig:cutbox}, with external momenta $K_r$, $r=1,2,3,4$. It corresponds to a diagram where the $D$-dimensional momenta $L_r$ running in the loop  are on-shell. Let us decompose each loop momenta into a four-dimensional part $l_r$ and an extra-dimensional part $l_{-2\epsilon}$, satisfying $l^2_{-2\epsilon}=-\mu^2$ (the same for all $L_r$ by momentum conservation, since the $K_r$ are four-dimensional). The cut conditions are
\begin{equation}
\label{cut1}
l^2=(l+K_1)^2=(l+K_1+K_2)^2=(l-K_4)^2=\mu^2\,.
\end{equation}
The cut conditions have two solutions, $l^\pm$, which depend on $\mu$. Each diagram gets a contribution from the two solutions, but we only pick up the coefficient of the leading power in $\mu$, which is $\mu^4$. 

We define
\begin{equation}
\label{badger}
C(i+1,\ldots, j|j+1,\ldots, k|k+1,\ldots, l|l+1,\ldots, i)=\frac{1}{2} \sum_{l^\pm} A_1\,A_2\,A_3\,A_4\,\Big|_{\mu^4} \,.
\end{equation}
The $A_r$ are the four subamplitudes of adjacent external gluons with a very massive scalar running in the loop. For instance, labelling the external gluons that go into $K_1$ as $i+1,\ldots,j$, we have
\begin{equation}
A_1 = A(l; i+1,\ldots, j; -l +K_1 ) \,,
\end{equation}
where $K_1=k_{i+1} + \ldots + k_j$. The full partial amplitude is given by
\begin{multline}
(4\pi)^2 A_{n;1}^{(1)}(1,2,\ldots,n) = \\
-\frac{1}{24}\sum_{i=1}^{n} \sum_{j=i+1}^{i-3} \sum_{k=j+1}^{i-2} \sum_{l=k+1}^{i-1} C(i+1,\ldots, j|j+1,\ldots, k|k+1,\ldots, l|l+1,\ldots, i) \,,
\end{multline}
using the notation $k_{n+i}\equiv k_i$. Recall that in Section~\ref{subsec:examples}, we explicitly confirmed this prescription in the particular case of the four-point amplitude, since we have shown that only the coefficient of $\mu^4$ contributed.

We can simplify the prescription of \cite{Badger:2008cm}, given in \eqref{badger}, by noting that we can determine explicitly the leading $\mu$ behaviour of the solutions $l^\pm$. Let us rewrite the cut conditions \eqref{cut1} as
\begin{equation}
\label{cut2}
l^2=\mu^2\,, \qquad 2l\cdot K_1 + K_1^2=2l\cdot (K_1+K_2) + (K_1+K_2)^2=-2l\cdot K_4 + K_4^2=0\,.
\end{equation}
We have one quadratic equation and three linear ones, so it is clear that there are only two solutions. We can use the external momenta to form a basis,
\begin{equation}
l^\lambda  = \alpha_1\, K_1^\lambda  + \alpha_2\, K_2^\lambda  + \alpha_3\, K_3^\lambda  + \alpha_\omega\, \omega^\lambda, \qquad \omega^\lambda= \epsilon^{\lambda \nu\rho\sigma} K_{1\nu}K_{2\rho}K_{3\sigma}\,,
\end{equation}
where $\epsilon^{\mu\nu\rho\sigma}$ is the Levi-Civita symbol. Now, in the linear equations of \eqref{cut2}, $l$ only appears contracted with momenta $K_r$. Therefore, the coefficient $\alpha_\omega$ does not appear in these equations, and they give a solution for $\alpha_1$, $\alpha_2$ and $\alpha_3$ which is independent of $\mu$. The coefficient $\alpha_\omega$ can then be determined using the quadratic equation, $l^2=\mu^2$, giving
\begin{equation}
\label{eq:llmu}
{l^\pm} = \pm \mu\, \bar{l} + {\mathcal O}(\mu^0)\,,
\end{equation}
where
\begin{equation}
\bar{l}^\lambda = \frac{\omega^\lambda}{\left( \omega \cdot \omega \right)^{1/2}}\,.
\end{equation}
For each subamplitude in \eqref{badger}, we only take the leading coefficient proportional to $\mu$. Moreover, each subamplitude is sensitive to the sign $\pm$ in \eqref{eq:llmu}, but not the product of the four subamplitudes. Therefore, we can substitute \eqref{badger} by 
\begin{equation}
\label{badger2}
C(i+1,\ldots, j|j+1,\ldots, k|k+1,\ldots, l|l+1,\ldots, i)= \bar{A}_1\,\bar{A}_2\,\bar{A}_3\,\bar{A}_4 \,,
\end{equation}
where we defined
\begin{equation}
\bar{A}_r=A_r(l=\bar{l})|_{\mu} \,.
\end{equation}
The subscript means that we take only the contribution linear in $\mu$. The subamplitudes $\bar{A}_r$ are just multiplied together, and the loop momentum running in the loop is the same in every subamplitude, so we conclude that each box is completely symmetric for the permutation of its corner subamplitudes. This will be crucial in the following.

Let us start with the four-point case. We have
\begin{equation}
(4\pi)^2 A_{4;1}^{(1)}(1,2,3,4) = - \frac{1}{6} C(1|2|3|4) \; \longrightarrow \; {\mathcal D}^{k_1k_2k_3k_4} \,.
\end{equation}
We want to identify the amplitude itself with ${\mathcal D}^{k_1k_2k_3k_4}$ (possibly up to factors) since it is symmetric for the permutation of the external legs. Indeed, each subamplitude $\bar{A}_r$ (associated with the external momentum $k_r$) is given by
\begin{equation}
\label{eq:subamp3pt}
\bar{A}_r =  \frac{e_r^{(+)}}{r_\eta} X(\bar{l}, r) = -\frac{1}{\langle\eta r \rangle^2} X(\bar{l}, r)\,.
\end{equation}
Notice that each subamplitude is {\it independently} invariant for the choice of the reference spinor $|\eta\rangle$. That is to say, we could choose a different spinor $|\eta_r\rangle$ for each subamplitude; the vertex factor $X_r$ will then depend on that choice,
\begin{equation}
X_r(i,j) = \langle \eta_r |ij| \eta_r \rangle \,.
\end{equation}
We can now write
\begin{equation}
\label{eq:Dkabcd}
 {\mathcal D}^{k_1k_2k_3k_4}_{1|2|3|4} = \frac{X_1(\bar{l}, 1)}{\langle\eta_1 1 \rangle^2} \, \frac{X_2(\bar{l}, 2)}{\langle\eta_2 2 \rangle^2}\, \frac{X_3(\bar{l}, 3)}{\langle\eta_3 3 \rangle^2} \, \frac{X_4(\bar{l}, 4)}{\langle\eta_4 4 \rangle^2} \,,
\end{equation}
so that
\begin{equation}
C(1|2|3|4) =  {\mathcal D}^{k_1k_2k_3k_4}_{1|2|3|4} \,.
\end{equation}
The definition \eqref{eq:Dkabcd} can be directly extended to higher points,
\begin{equation}
\label{eq:DKabcd}
 {\mathcal D}^{K_1K_2K_3K_4}_{i+1,\ldots, j|j+1,\ldots, k|k+1,\ldots, l|l+1,\ldots, i} = 
\frac{X_1(\bar{l}, K_1)}{\alpha_1^{(i+1,\ldots, j)}} \,
\frac{X_2(\bar{l}, K_2)}{\alpha_2^{(j+1,\ldots, k)}} \,
\frac{X_3(\bar{l}, K_3)}{\alpha_3^{(k+1,\ldots, l)}} \,
\frac{X_4(\bar{l}, K_4)}{\alpha_4^{(l+1,\ldots, i)}} \,,
\end{equation}
where the momenta $K_r$ are the overall momenta entering each of the four subamplitudes, and where the external factors are given by
\begin{equation}
\alpha_r^{(i+1,\ldots, j)} = \prod_{s=i+1}^j \left( -\langle\eta_r s \rangle^2 \right)\,.
\end{equation}

\begin{figure}
\centering
\includegraphics[scale=0.5]{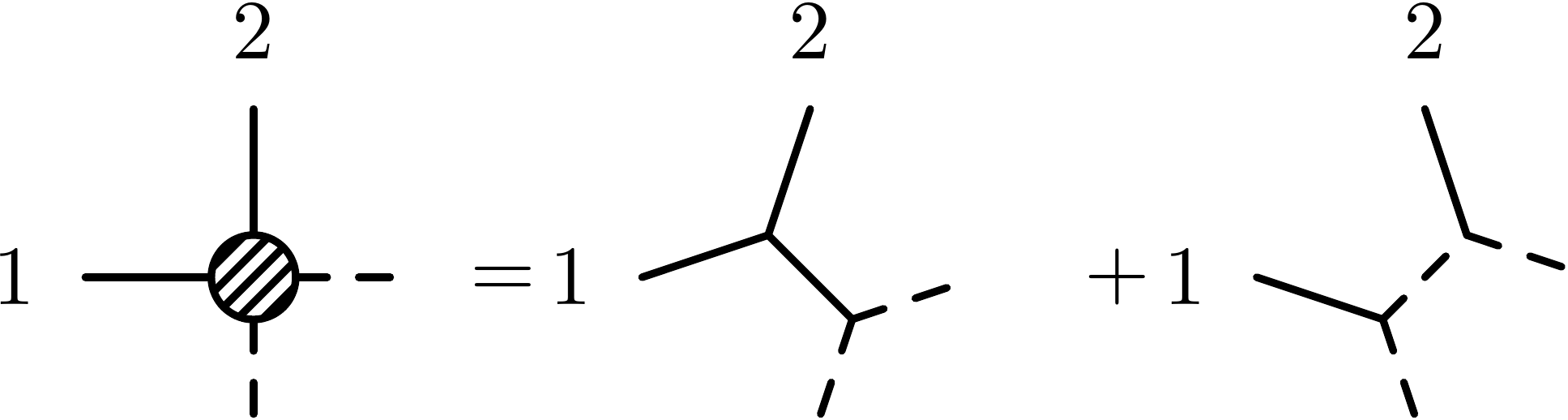}
\caption{This subamplitude is part of the pentagon contribution to the box cut.}
\label{fig:pentCorner}
\end{figure}

Let us now consider the five-point case,
\begin{multline}
\label{eq:A5C}
(4 \pi)^2 A_{5;1}^{(1)}(1,2,3,4,5) = -\frac 16 \bigg( C(12|3|4|5) + C(51|2|3|4) \\
+ C(45|1|2|3) + C(34|5|1|2) + C(23|4|5|1) \bigg) .
\end{multline}
Based on \eqref{eq:5ptsFD}, we would like to make the identification
\begin{equation}
\label{eq:5ptvertices}
C(12|3|4|5) \; \longrightarrow \; {\mathcal F}^{k_1k_2(-q)}\frac{i}{s_{12}} {\mathcal D}^{qk_3k_4k_5} \,.
\end{equation}
Using the prescription \eqref{badger2}, we have that the associated subamplitudes $\bar{A}_2$, $\bar{A}_3$ and $\bar{A}_4$ are given as in \eqref{eq:subamp3pt}, with the appropriate external momentum. However, the subamplitude $\bar{A}_1$ has two external gluons, and is given by
\begin{equation}
\label{eq:subamp4pt}
\bar{A}_1 = \frac{i}{\langle\eta_1 1 \rangle^2 \langle\eta_1 2 \rangle^2}
\left( 
\frac{X_1(1,2)X_1(\bar{l},1+2)}{s_{12}} +
\frac{X_1(\bar{l}, 1)X_1(\bar{l}+1,2)}{2 \bar{l}\cdot k_1} 
\right) \Bigg|_\mu \,.
\end{equation}
The second term in this expression leads a pentagon-like structure in $C(12|3|4|5)$, since all five external gluons connect directly with the massive scalar in the loop; see Figure~\ref{fig:pentCorner}. The identification \eqref{eq:5ptvertices} is clearer in a gauge where that term vanishes, so that we can factorize the contributions from the ``vertices" ${\mathcal F}$ and ${\mathcal D}$. One such gauge is
\begin{equation}
|\eta_1 \rangle \quad \to \quad   \bar{l}|1\rangle \,,
\end{equation}
so that $X_1(\bar{l}, 1)=0$. We can now make the identifications
\begin{align}
{\mathcal D}^{K_1k_3k_4k_5}_{12|3|4|5} &= - \frac{X_1(\bar{l}, 1+2)}{\langle\eta_1 1 \rangle^2\langle\eta_1 2 \rangle^2}\, \frac{X_2(\bar{l}, 3)}{\langle\eta_2 3 \rangle^2}\,
\frac{X_3(\bar{l}, 4)}{\langle\eta_3 4 \rangle^2}\,
\frac{X_4(\bar{l}, 5)}{\langle\eta_4 5 \rangle^2} \,, \\
{\mathcal F}^{k_1k_2 (-K_1)}_1 &= X_1(1,2) \,, \qquad \quad K_1 = k_1+k_2\,,
\end{align}
so that
\begin{equation}
\label{eq:C5FD}
C(12|3|4|5) = {\mathcal F}^{k_1k_2(-K_1)}_1 \frac{i}{s_{12}} {\mathcal D}^{K_1k_3k_4k_5}_{12|3|4|5}\,.
\end{equation}
Using the kinematic structure constants, we obtained a representation of the five-point amplitude which makes manifest the ``vertex" structure in \eqref{eq:5ptsFD}. There are several gauge choices involved, as each term in \eqref{eq:5ptsFD} -- or, equivalently, each box contribution $C$ in \eqref{eq:A5C} -- requires a different gauge choice.

The same pattern repeats at higher points. For any subamplitude $\bar{A}_r$ of a contribution $C$ to the amplitude, one can choose a reference spinor $|\eta_r\rangle$ which eliminates all diagrams in the subamplitude which are not box-like (such as the pentagon-like example above). Notice that a subamplitude is invariant for the scaling of $|\eta_r\rangle$, so that we can choose $|\eta_r\rangle=(1\;\; x)$. The terms to be eliminated only depend on $|\eta_r\rangle$ in the numerator, and through $X_r$, so that we have a polynomial equation for the gauge parameter $x$.\footnote{We have explicitly checked that non-singular solutions exist up to seven points.} Once all subamplitudes are written in an appropriate gauge, the factor $X_r(\bar{l},K_r)$ can be absorbed into the four-point ``vertex" ${\mathcal D}$, together with the external particle factors, while all the other $X_r$ correspond to ``vertices" ${\mathcal F}$.

Let us point out that there is an analogous structure described recently in superstring amplitudes \cite{Mafra:2012kh}. At one loop, in the field theory limit, open superstring amplitudes are related to the tree-level $F^4$ contribution at order $\alpha'^2$. This contribution is in turn related to the all-plus one-loop amplitudes under study here, as first discussed in \cite{Stieberger:2006bh,Stieberger:2006te}.

Ref.~\cite{BjerrumBohr:2011xe} also analysed relations between one-loop one-minus amplitudes, and it would be interesting to investigate that case along the lines followed here. The method for computing rational terms presented in \cite{Badger:2008cm} can also be applied, but it requires the inclusion of triangles and bubbles.

%%%%%%%%%%%%%%%%%%%%%%%%%%%%%%%%%%%%%
%%%%%%%%%%%%%%%%%%%%%%%%%%%%%%%%%%%%%

\section{A series of colour-dual form factors}
\label{sec:ffs}

In this section it is shown that the form factor of the anti-self-dual Lagrangian with all-plus helicity gluons,
\be \langle \textrm{tr} (F_-^2)(x) | + + \ldots + \rangle \label{eq:defff} \ee
admits an explicitly colour dual perturbation theory at tree level. This follows by calculating this form factor using the self-dual Yang-Mills theory. As a side product our calculation shows that this particular theory admits an infinite series of observables of which the form factor in equation \eqref{eq:defff} is an example.

The first step is to Fourier transform the operator in the form factor in equation \eqref{eq:defff} to momentum space. Its momentum is denoted $q$. Before starting any calculation it should be noted this particular form factor has a known expression at tree level from its relation to effective Higgs-gluon couplings, see \cite{Dixon:2004za},
\be \langle \textrm{tr} (F_-^2)(x) | + + \ldots + \rangle = \frac{(q^2)^2}{\sbraket{12}\sbraket{23} \ldots \sbraket{n1}}\ee
where colour-ordered gluons are labelled $1$ through $n$. This form can be understood from the collinear factorization properties of like-helicity gluons and the symmetry properties.

The argument that the above form factor has an explicitly colour-dual representation follows from embedding the form factor calculation into the self-dual Yang-Mills theory. The starting point for this is the Chalmers and Siegel action for full Yang-Mills theory \cite{Chalmers:1996rq},
\be S = \int d^4x \textrm{tr} \left( B F^+ - \frac{1}{2} B^2\right) \ee
Integrating out the field $B$ yields the usual Yang-Mills theory (up to a topological term). Dropping the $B^2$ term gives self-dual Yang-Mills theory, which will be done from now on. The field equation for $B$ in this case sets $F^+$ to vanish. To fix conventions set
\be F^+_{\dot{\alpha} \dot{\beta}} = F_{\alpha \dot{\alpha} \beta \dot{\beta}} \epsilon^{\alpha \beta} \ee
and
\be F^-_{\alpha \beta} = F_{\alpha \dot{\alpha} \beta \dot{\beta}} \epsilon^{\dot{\alpha} \dot{\beta}} \ee
light-cone gauge is given as before by
\be \eta_{\alpha}\eta_{\dot{\alpha}} A^{\alpha \dot{\alpha}} = 0 \ee
The spinorial form of the light-cone condition has two natural solutions,
\be A^{\alpha \dot{\alpha}} \propto \eta^{\alpha} A^{\dot{\alpha}} \qquad \textrm{or} \qquad  A^{\alpha \dot{\alpha}} \propto \eta^{\dot{\alpha}} A^{\alpha}  \ee
Note that the first make the interaction term in $F^+$ vanish, while the second does the same for $F^-$. Let us pick the second solution. A complete basis for the spinor space is spanned by
\be \{\eta_{\alpha}, \tilde{\eta}_{\alpha} ,\eta_{\dot{\alpha}}, \tilde{\eta}_{\dot{\alpha}}  \} \ee
where $\braket{\tilde{\eta} \eta} = 1 = \sbraket{\tilde{\eta} \eta}$. In terms of these spinors one can decompose the symmetric tensor $B_{\dot{\alpha} \dot{\beta}}$ as
\be B_{\dot{\alpha} \dot{\beta}} = B \tilde{\eta}_{\dot{\alpha}} \tilde{\eta}_{\dot{\beta}}  +B' \left( \eta_{\dot{\alpha}} \tilde{\eta}_{\dot{\beta}}  + \eta_{\dot{\beta}} \tilde{\eta}_{\dot{\alpha}}  \right) + B'' \eta_{\dot{\alpha}} \eta_{\dot{\beta}} \ee
Plugging this decomposition into the Lagrangian in the chosen light-cone gauge gives
\begin{equation} \textrm{tr} \left( B F^+ \right) =  \textrm{tr} \left( B  \tilde{\eta}_{\dot{\alpha}}  \tilde{\eta}_{\dot{\beta}} F^{+,\dot{\alpha} \dot{\beta}}  + B'  \left( \eta_{\dot{\alpha}} \tilde{\eta}_{\dot{\beta}}  + \eta_{\dot{\beta}} \tilde{\eta}_{\dot{\alpha}}  \right)   F^{+,\dot{\alpha} \dot{\beta}}\right)\end{equation}
where the last term drops out in this gauge. Furthermore,
\be \left( \eta_{\dot{\alpha}}  \tilde{\eta}_{\dot{\beta}}  + \eta_{\dot{\beta}}  \tilde{\eta}_{\dot{\alpha}}  \right) F^{+,\dot{\alpha}\dot{\beta}} \propto  \eta_{\dot{\alpha}} p^{\dot{\alpha} \alpha} A_{\alpha} \ee
so that integrating out $B'$ yields
\be A_{\alpha} = \eta_{\dot{\alpha}} p^{\dot{\alpha}}{}_{\alpha}  A \ee
Plugging this back into the action and collecting gives
\be \mathcal{L} = \textrm{tr} \left( B F^+ \right) = -B \Box A + B \eta^{\dot{\alpha}}  \eta^{\dot{\beta}}  (\partial_{\alpha \dot{\alpha}} A) (\partial^{\alpha}{}_{\dot{\beta}} A) \ee
Which is the usual self-dual Yang-Mills theory in light-cone gauge. Important is that introducing a $J \textrm{tr}  (F^-)^2$ current term into the action does not change the derivation. In the particular gauge under study the operator  $(F^-)^2$ can be expressed in terms of the fields as
\begin{equation}\label{eq:antiselfdualop}
\textrm{tr} (F^-)^2 = \textrm{tr}   \left( \eta^{\dot{\alpha}} \eta^{\dot{\beta}}  (\partial_{\alpha \dot{\alpha}} \partial_{\beta,\dot{\beta}}  A) (  \eta_{\dot{\delta}} \eta_{\dot{\gamma}}  (\partial^{\alpha \dot{\delta}} \partial^{\beta,\dot{\gamma}}A) \right)
\end{equation}
A form factor of this operator and an arbitrary amount of like helicity fields can therefore be calculated purely in self-dual Yang-Mills theory. As an example, consider
\be \langle \textrm{tr} (F^-)^2 | + +  \rangle \ee
at tree level. This can simply be calculated by putting the fields on-shell in \eqref{eq:antiselfdualop}. This gives
\be \langle \textrm{tr} (F^-)^2 | + +  \rangle = \braket{12}^2   = \frac{(q^2)^2}{\sbraket{12}^2}\ee
as it should. This shows this form factor can be computed with an explicitly colour-dual perturbation theory. This computation can be extended to form-factors with multiple insertions of the field strength tensor, generalizing \eqref{eq:defff}. For instance, one could consider
\be \langle \left( \textrm{tr} (F_-^2)(x)\right)\left( \textrm{tr} (F_-^2)(x)\right) | + + \ldots + \rangle  \ee
In the self-dual sector, this form factor is structurally simply the product of two of the MHV form factors. The only complication is the fact that this is a multi-trace object, so color-ordering has to be defined with respect to two traces. Generalizing to more inserted operators of this type is straightforward.

%%%%%%%%%%%%%%%%%%%%%%%%%%%%%%%%%%%%%
%%%%%%%%%%%%%%%%%%%%%%%%%%%%%%%%%%%%%

\section{Discussion and conclusions}
\label{sec:dc}
Above we have obtained the first series of examples of colour-dual numerators at any loop level to all multiplicity. These series follow by extension of the observation of \cite{Monteiro:2011pc} that self-dual Yang-Mills theory obeys colour-kinematics duality at the level of the Lagrangian. Since this theory generates the integrand of the one-loop helicity-equal amplitudes, this integrand is obtained in manifestly colour-dual form. Moreover, by exploiting gauge freedom the same results can be obtained for the one-minus helicity integrand. Interestingly, these two series of examples comprise all known finite amplitudes in Yang-Mills theory. We have checked explicitly in the four particle case that the integrands of the gauge theory amplitudes indeed integrate to the known results. Moreover, we have shown that the expressions obtained by double copy also integrate to the correct results for four points in the corresponding gravitational theory. It would be interesting to explore in more detail how colour-kinematics duality could be used to simplify the form of these integrands. The role of generalised gauge transformations should receive special attention in this exploration. 

It has also been shown above that the integrated expressions for the all-plus amplitudes have a residual colour-kinematics interpretation. This can be exposed for each separate massive box coefficients by choosing a special gauge. The structure thus found here has been used in a conjectural form in \cite{BjerrumBohr:2011xe} to inspire certain relations between one-loop all-plus amplitudes. By the known fact that these relations extend to massive box coefficients regardless of helicity \cite{Boels:2011tp} it is easy to speculate a similar colour-kinematics interpretation must exist in this more general case. Investigating this should prove useful. 

It would be interesting to obtain an explicitly colour-dual form of the gauge theory Lagrangian beyond the self-dual truncation, along the lines of~\cite{Bern:2010yg}. Extending this to the full gravitational Lagrangian would be the logical next step. Of course, the existence of these Lagrangians in a general form would prove colour-kinematics duality. Moreover, the duality would apply to observables calculated in these theories, such as correlation functions and form factors. In the latter case we have shown that the self-dual Lagrangian can yield some insight already. As far as we know the observation that this theory has any tree level gauge-invariant observables beyond a single three point amplitude constitutes a new result. Perhaps this can yield some inspiration to find proper observables in (0,2) theories in six dimensions beyond amplitudes~\cite{Huang:2010rn}.

Our results in non-supersymmetric Yang-Mills and gravity theory at one loop may have interesting implications for maximally supersymmetric Yang-Mills ($\mathcal{N}=4$) and supergravity ($\mathcal{N}=8$). A conjecture made in \cite{Bern:1996ja} relates the integrand of the helicity equal amplitudes in non-supersymmetric Yang-Mills directly to the integrand of the maximally helicity violating (MHV) amplitude in $\mathcal{N}=4$ Yang-Mills. A similar conjecture was made in \cite{Bern:1998sv} for the relation between helicity equal amplitudes in Einstein gravity and $\mathcal{N}=8$ supergravity. Taken together with these conjectures, our results suggest that colour-kinematics duality holds for MHV amplitudes in $\mathcal{N}=4$ Yang-Mills and that the double copy construction yields the correct $\mathcal{N}=8$ supergravity result for this class of amplitudes.

\appendix

\acknowledgments
We thank Emil Bjerrum-Bohr, John Joseph Carrasco, Tristan Dennen, Henrik Johansson, Oliver Schlotterer, Gang Yang and especially Simon Badger for helpful conversations during the course of our work. We also thank the anonymous referee for useful suggestions. RB and RSI would like to thank the Niels Bohr International Academy for hospitality during the completion of this work. Likewise, RM and DOC would like to thank the University of Hamburg for hospitality. This work has been supported by the German Science Foundation (DFG) within the Collaborative Research Center 676 ``Particles, Strings and the Early Universe".

\bibliographystyle{jhep}

\end{document}